\documentclass[fleqn,usenatbib]{mnras}

\usepackage{newtxtext,newtxmath}

\usepackage[T1]{fontenc}

\DeclareRobustCommand{\VAN}[3]{#2}
\let\VANthebibliography\thebibliography
\def\thebibliography{\DeclareRobustCommand{\VAN}[3]{##3}\VANthebibliography}

\usepackage{graphicx}	
\usepackage{amsmath}	

\newcommand{\ncandidates}{878}
\newcommand{\nrgb}{6\,878\,665}

\title[GC-Origin Stars from Gaia BP/RP Spectra]{The ones that got away: chemical tagging of globular cluster-origin stars with Gaia BP/RP spectra}

\author[S. G. Kane et al.]{
Sarah G. Kane,$^{1}$\thanks{E-mail: sgk27@cam.ac.uk (SGK)}
Vasily Belokurov,$^{1}$ Miles Cranmer,$^{1,2,3}$ Stephanie Monty,$^{1}$ Hanyuan Zhang,$^{1}$ \newauthor Anke Ardern-Arentsen,$^{1}$ and Elana Kane$^{1}$
\\
$^{1}$Institute of Astronomy, University of Cambridge, Madingley Road, Cambridge CB3 0HA, UK\\
$^{2}$Department of Applied Mathematics and Theoretical Physics, University of Cambridge, Wilberforce Road, Cambridge CB3 0WA, UK\\
$^{3}$Kavli Institute for Cosmology, Cambridge, University of Cambridge, Madingley Road, Cambridge CB3 0HA, UK
}

\date{Accepted XXX. Received YYY; in original form ZZZ}

\pubyear{\the\year{}}

\begin{document}
\label{firstpage}
\pagerange{\pageref{firstpage}--\pageref{lastpage}}
\maketitle

\begin{abstract}
Globular clusters (GCs) are sites of extremely efficient star formation, and recent studies suggest they significantly contributed to the early Milky Way's stellar mass build-up. Although their role has since diminished, GCs' impact on the Galaxy's initial evolution can be traced today by identifying their most chemically unique stars—those with anomalous nitrogen and aluminum overabundances and oxygen depletion. While they are a perfect tracer of clusters, be it intact or fully dissolved, these high-[N/O], high-[Al/Fe] GC-origin stars are extremely rare within the current Galaxy.
To address the scarcity of these unusual, precious former GC members, we train a neural network (NN) to identify high-[N/O], high-[Al/Fe] stars using low-resolution \textit{Gaia} BP/RP spectra. Our NN achieves a classification accuracy of approximately $\approx99\%$ and a false positive rate of around $\approx7\%$, identifying \ncandidates~new candidates in the Galactic field. We validate our results with several physically-motivated sanity checks, showing, for example, that the incidence of selected stars in Galactic GCs is significantly higher than in the field. Moreover, we find that most of our GC-origin candidates reside in the inner Galaxy, having likely formed in the proto-Milky Way, consistent with previous research. The fraction of GC candidates in the field drops at a metallicity of [Fe/H]$\approx-1$, approximately coinciding with the completion of \textit{spin-up}, i.e. the formation of the Galactic stellar disk.
\end{abstract}




\section{Introduction}

Globular clusters (GCs) are sites of extremely efficient star formation. Comprised of $10^5-10^7$ of densely-packed stars that form at approximately the same time, these objects are suggested to the primary contributors to galaxy formation at early times \citep[e.g. as nuclear star clusters, building blocks of stellar halos, and of chemically diverse discs and spheroids through their association with star forming clumps;][]{tremaine_1975, gnedin_2014, Martell_2011, Schiavon_2017, belokurov_kravstov_nitrogen, Clarke_2019, Debattista_2023}. Observations at high redshift are now beginning to suggest that this may indeed be the case. For instance, \citet{mowla_firefly_sparkle} have identified in the JWST imaging a $z=8.3$ galaxy with approximately half of its mass locked in massive star-forming clusters. The interpretation of these results is of ongoing community interest, with \citet{Rusta_2024_firefly_interpretation} suggesting that these bright, unresolved stellar clumps may instead be "building block galaxies" themselves.

Nearby GCs, resolved into individual stars, ubiquitously exhibit a light-element anti-correlation indicative of multiple stellar populations \citep[early evidence of this phenomena emerged through measurements of the molecule CN in MW GCs, e.g.][]{Freeman_1975, Norris_1975, Bessel_1976, Cottrell_1981}. Stars in the first population (1P), so called because they are thought to have formed first, display light element abundances consistent with MW field stars at similar metallicities. By contrast, members of the second population (2P) are exceptionally enhanced in nitrogen, aluminum, and sodium and depleted in oxygen, carbon, and magnesium relative to field stars \citep{Gratton_2004,Carretta_2009a,Carretta_2009b,Carretta2010, milone_multiple_pops_GCs}.
The precise origin of the anomalous chemistry of the 2P stars in GCs is not fully understood, with theories invoking enrichment from supermassive stars \citep{Denissenkov_supermassive_stars_GCs, Gieles_2018} among the 1P, winds from AGB stars \citep[e.g.][]{Ventura_2009, DAntona_2016} in the dense cluster environment and massive binary interactions \citep[for a more complete summary, see][]{Bastian_multiple_pops_in_GC,Gratton_what_is_GC,milone_multiple_pops_GCs}.
Nonetheless, the light-element anticorrelation is distinctive to and universal across GCs \citep[][]{Bastian_multiple_pops_in_GC,Gratton_what_is_GC,milone_multiple_pops_GCs}. 
{By comparison, within many clusters members exhibit relatively small spreads in metallicity on the order of about 0.1~dex among 1P stars and smaller still among 2P stars \citep{Marino_2019,Marino_2023,Legnardi_2022}. Notably, however, some anomalous clusters (Type II GCs; ex., $\omega$~Centauri) have much
more substantial metallicity variations across all of their stellar populations \citep{Marino_2015,milone_multiple_pops_GCs}.

The anomalous chemistry of the 2P GC stars is so distinct relative to MW field stars that we can tag field stars with this pattern of overabundances and depletions as having originated in a GC; nitrogen overabundance in particular has been used consistently for this form of chemical tagging \citep{Horta_nrich_stars, Schiavon_2017, belokurov_kravstov_nitrogen,Martell_2010,Phillips_2022}. \citet{belokurov_kravstov_nitrogen} use APOGEE data \citep[][]{APOGEE_DR17} to identify MW field stars with high [N/O] ratio and show that the majority of these likely GC 2P members are part of \textit{Aurora}, the old ([Fe/H]$\lesssim-1.3$), \textit{in-situ} MW halo \citep{belokurov_aurora,rix_poor_old_heart}. Moreover, \citet{belokurov_kravstov_nitrogen} find that the fraction of high-[N/O] field stars in the Galactic halo drops rapidly with increasing metallicity around [Fe/H]~$\simeq-1.0$. This metallicity dependence of the high-[N/O] fraction indicates that in the early MW a significant portion of star formation was locked in GCs but the clusters' relative contribution to the Galaxy's stellar mass buildup began to decline around the time of the emergence of the MW thick disk, or at the time of “spin-up” \citep[see also][]{belokurov_aurora, chandra_three_phase,Belokurov_2024_GCs, Zhang2024}. In spite of their hypothesized importance for tracing the early history of the Galaxy, high-[N/O] stars are exceptionally rare in the present-day Galactic field, constituting only $2-3\%$ of the inner halo  and even lower further out from the galactic centre \citep{Martell_2016,Horta_nrich_stars,Schiavon_2017, belokurov_kravstov_nitrogen}. The scarcity of these chemically anomalous stars poses a real challenge for investigating the history of GCs in the MW.

The anomalous nitrogen enhancement of 2P GC stars has become particularly relevant within the context extragalactic astronomy given the recently discovered nitrogen-rich chemistry of GN-z11, a $z=10.6$ galaxy discovered first with Hubble and then observed with JWST \citep{bunker_gnz11,Cameron_nitrogen_gnz11}. The extreme nitrogen enrichment of GN-z11 and in several other galaxies at high redshift \citep[e.g.][]{Ji2024_nitrogen_loud, Topping2024_nitrogen, Yanagisawa2024_nitrogen} along with the mounting evidence that GCs contribute significantly to star formation in the early MW \citep[][]{belokurov_kravstov_nitrogen} has spurred theories that the source of nitrogen enrichment in 2P GC stars and in high-z Universe may be one and the same. One of the more popular hypotheses is that in GN-z11 supermassive stars are responsible for the elevated N abundance \citep{Charbonnel_gnz11_supermassive_stars_nitrogen,Denissenkov_supermassive_stars_GCs, nandal_supermassive_stars_nitrogen}. Alternative theories have also been proposed, such as an intermittent star formation history \citep{kobayashi_gnz11}. Nonetheless, this is an active area of research that makes the study of clusters extremely relevant.

The scarcity of high-[N/O] stars in the Galactic field poses a challenge for using these stars as a tracer of the history of globular clusters in the MW. High resolution spectroscopic surveys are both limited in how many stars they can observe as well as the difficulty in measuring N at all wavelengths, thus making the number of known 2P GC-type stars outside of clusters limited. The European Space Agency's \textit{Gaia} mission \citep{gaia_mission} has recently revolutionized our understanding of the Milky Way by providing astronometric, photometric, and spectroscopic data for hundreds of millions of stars. Among the products from the most recent data release (Data Release 3; DR3) are low-resolution (R$\approx$30-100) Blue Photometer/Red Photometer (BP/RP) spectra for approximately 220 million stars \citep{gaia_dr3, gaia_bprp_processing, Gaia_bprp_calibration}. Combined, the full BP/RP spectra have a resolution of $R\sim60$ and cover wavelengths spanning from 330~nm to 1050~nm, which includes a CN band at 388~nm, a second CN feature at 421.5~nm for stars with [Fe/H]$\sim$-1, and a NH band at the edge of the wavelength range at 336~nm. Usefully, the resolution of the Blue Photometer is highest at lower wavelengths, where the nitrogen features are located \citep{Carrasco_2021_gaia_bprp}. Thus, we propose to leverage the large quantity of low resolution \textit{Gaia} BP/RP spectra to identify new high-[N/O] candidates.

The usefulness of the \textit{Gaia} BP/RP spectra to derive various stellar parameters and abundances has been of broad interest to the community \citep[see e.g.][]{witten_information_bprp}, and a wealth of papers on the topic have been published in the past few years. Many of these recent works rely on machine learning methods to extract information from the BP/RP spectra, as the low resolution nature of the data makes more traditional abundance line analysis challenging. Among these works, multiple groups have derived stellar effective temperatures, surface gravities, and metallicities from the BP/RP spectra with relatively high accuracy \citep[e.g.][]{andrae_metallicities, Liu_bprp, Yao2024_bprp, Khalatyan2024_bprp}. Other works have extracted $\alpha$-element abundances from the spectra as well as the three fundamental stellar parameters \citep[e.g.][]{li_alpha_bprp, Hattori_2024}. Some groups have also used the BP/RP spectra to identify chemically peculiar stars. For instance, \citet{lucey_cemp} identified carbon-enhanced metal-poor star candidates from the BP/RP spectra; while \citet{Sanders2023_crich} used BP/RP to pick out C-rich long periodic variables. Recently, \citet{fallows_sanders_bprp} used a neural network to derive data-driven $T_{\textrm{eff}}$, $\log g$, metallicity, [$\alpha$/Fe], [C/Fe], and [N/Fe] abundances from the BP/RP spectra.

In this work, we leverage the vast quantity of \textit{Gaia} BP/RP spectra to detect new high-[N/O] stars in the MW halo. Thanks to {\it Gaia}, we are able to increase the number of known halo stars with 2P-type chemistry by a factor of as compared to APOGEE. We perform heteroscedastic regression with a simple neural network to derive our own estimates of $T_{\textrm{eff}}$, $\log g$, [Fe/H], [N/O] and [Al/Fe] from the BP/RP spectra, and from these predictions, we classify stars as high-[N/O], high-[Al/Fe] candidates. With these classifications, we identify \ncandidates~new high-[N/O] candidates in the MW halo. Using these new high-[N/O] field stars, we are able to use their properties and distribution to study the contribution of GCs to the Galaxy at an unprecedented scope. 

Our paper is organised as follows. Section \ref{sec:methods} describes our data-driven approach to locating new high-[N/O] candidates in three subsections. In Section \ref{sec:data}, we discuss our training dataset, and in Section \ref{sec:algorithm}, we outline the neural network we use to infer stellar parameters and [N/O] and [Al/Fe] abundances. Section \ref{sec:andrae_giants} details the sample of red giant stars upon which we use our trained neural network to identify new high-[N/O] stars. We validate the results of our network performance in Section \ref{sec:validation}. The rest of Section \ref{sec:results} details our newly identified high-[N/O] candidates, with Section \ref{sec:additional_tests} detailing additional tests we used to verify the reliability of our selection. Section \ref{sec:nrich_properties} discusses the properties of our candidates and their distribution within the Galaxy. We provide a summary of our analysis in Section \ref{sec:conclusions}.

\section{Data and Methods}
\label{sec:methods}

\subsection{Training and Validation Data}
\label{sec:data}

\begin{figure*}
	\includegraphics[width=6.3in, alt={In the left, the CMD of stars in the training and validation data clearly populate only the red giant branch, whereas the comparison CMD extends into the main sequence. On the right, the high-[N/O] stars comprise only a relatively small fraction of the total training and validation data, with most of the high-[N/O] stars belonging to GCs.}]{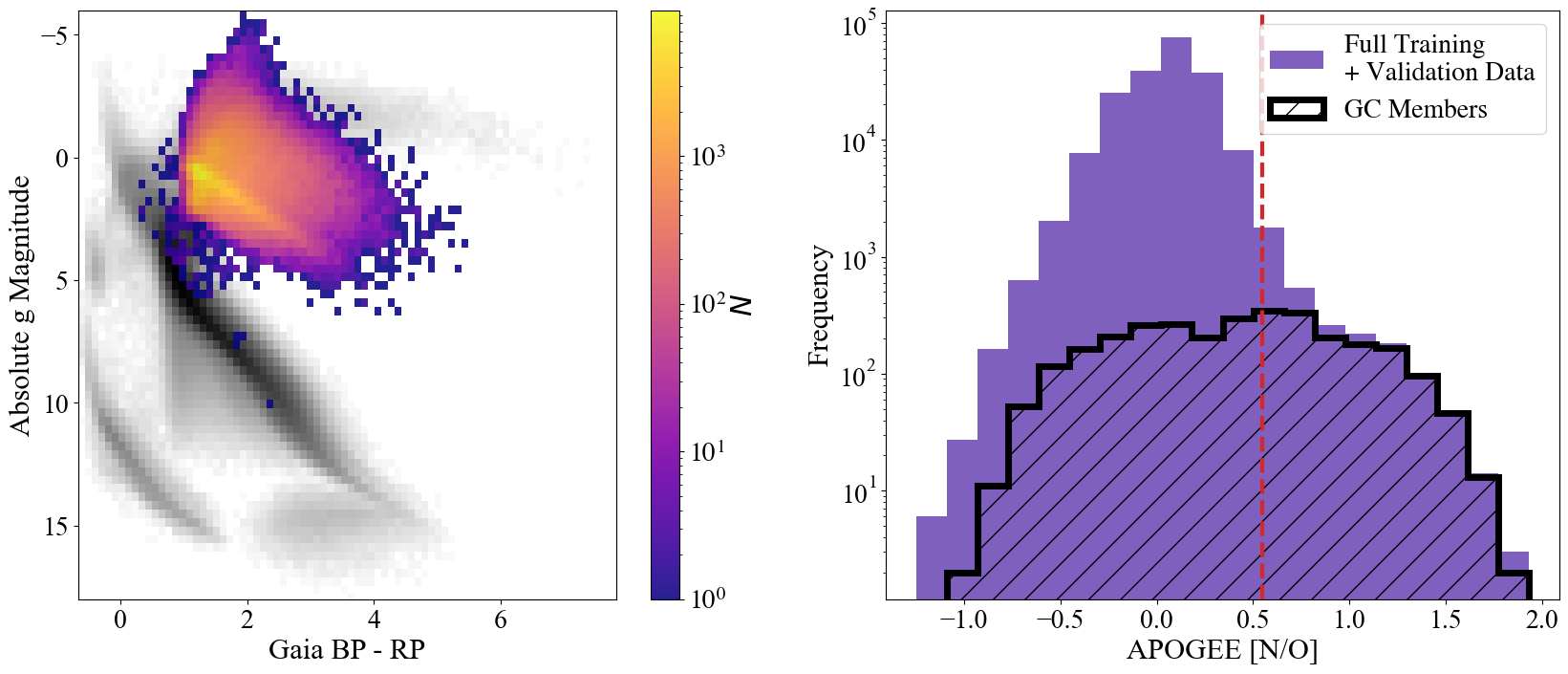}
    \caption{Left: The color-magnitude diagram (CMD) of our training and validation dataset of APOGEE and Gaia BP/RP sources (colored 2D histogram) overlaid on the CMD of 32\,459\,252 randomly selected BP/RP sources for comparison (grey 2D histogram). The extension of the RGB into redder colors is caused by dust effects. Right: The distribution of APOGEE [N/O] abundances in our training and validation dataset. The purple histogram shows the normalized distribution of [N/O] abundances in the dataset overall, and the transparent black histogram shows the distribution of [N/O] abundances of stars in our training and validation dataset tagged as GC members by \citet{vasiliev_gc}. The vertical, dashed red line marks the nominal cutoff of high-[N/O] stars are [N/O]$>$0.55. The GC members are included in the overall training and validation datasets; note that almost all high-[N/O] giants from APOGEE are tagged as GC members.}
    \label{fig:training_data}
\end{figure*}

Because our goal is to predict [N/O] and [Al/Fe] abundances from the \textit{Gaia} BP/RP coefficients, we require stars with both BP/RP observations in \textit{Gaia} DR3 and high-quality abundance measurements with which to train our neural network. To this effect, the core of our training data is comprised of cross-matches between \textit{Gaia} BP/RP observations and stars with data from the Apache Point Observatory Galactic Evolution Experiment \cite[APOGEE]{APOGEE_DR17}. APOGEE is unique among spectroscopic surveys and particularly well-suited to our goal of having training labels for [N/O] abundances because APOGEE spectra contain many N lines \citep[see][]{Schiavon_2017}. The robust nitrogen abundances provided by APOGEE have prompted its use in previous studies of nitrogen-rich stars \citep[e.g.,][]{Schiavon_2017, Horta_nrich_stars, belokurov_kravstov_nitrogen}, which has the added benefit of providing some consistency between our data and that of previous works. Despite known biases in the absolute value of O in APOGEE \citep{Carrillo_2022}, it has been shown to trace other $\alpha$ elements to high fidelity. From these cross-matches, we restrict our training data to giants by selecting stars with $T_\mathrm{eff}<5200$ and $\log g<3.0$. We further apply the following quality cuts to the abundances and stellar parameters from APOGEE:
\begin{itemize}
  \item Bitmasks to remove telluric contamination and duplicate targets (\texttt{EXTRATARG TELLURIC} \& \texttt{DUPLICATE})\footnote{\url{https://www.sdss4.org/dr17/irspec/apogee-bitmasks/}}
  \item \texttt{ASPCAP} flags: $\texttt{STAR\_BAD}$, $\texttt{TEFF\_BAD}$, \& $\texttt{LOGG\_BAD}<0$
  \item \texttt{SNR}~$>20$
  \item Errors for [Fe/H], [Mg/Fe], [N/Fe], [O/Fe], \& [Al/Fe]~$<0.1$
\end{itemize}
We also apply several cuts based on the \textit{Gaia} catalog to ensure the exclusion of potentially problematic BP/RP spectra from our training data.
\begin{itemize}
  \item $\texttt{phot\_g\_mean\_mag}<16.0$
  \item $\texttt{parallax}/\texttt{parallax\_error}>3.0$ \& $\texttt{parallax}>0$
  \item $\texttt{RUWE}<1.4$ \citep[to remove clear binary contamination per][]{belokurov_binaries}
\item $\texttt{bp\_chi\_squared}<10.5\times \texttt{bp\_degrees\_of\_freedom}$
\end{itemize}

After applying these cuts, our set of BP/RP-APOGEE cross-matches contains $199\,078$ stars, of which we use $80\%$ for training and withhold a randomly selected $20\%$ subset upon which to validate our network. The color-magnitude diagram (CMD) of our full training and validation dataset is shown in the left panel Fig.~\ref{fig:training_data}, wherein our selected stars clearly populate the red giant branch (RGB). The right panel of Fig.~\ref{fig:training_data} depicts the normalized distribution of [N/O] abundances in our training data and the distribution of [N/O] abundances from the subset of our training data that are identified as likely GC members in \citet{vasiliev_gc}. As is consistent with the presence of a chemically anomalous 2P population in GCs, the GC members have a much higher fraction of high-[N/O] stars as compared to the training dataset overall. Within the APOGEE data, we identify true high-[N/O] stars as those with $\textrm{[N/O]}-\textrm{[N/O] error}>0.55$ and [Al/Fe]~$>-0.1$. The cuts at [N/O]~$>0.55$ and [Al/Fe]~$>-0.1$ are consistent with the criteria used in \citet{belokurov_kravstov_nitrogen} to select a sample of 2P stars in the field without contamination (e.g., from nitrogen-rich but not oxygen-depleted or aluminum enhanced stars that thus do not display the full pattern of light element anticorrelations typical of 2P GC-origin stars). Using this selection, 1\,755 stars among the combined training and validation data are high-[N/O], high-[Al/Fe] stars.

The \textit{Gaia} BP/RP spectra are provided as a set of 110 coefficients of a set of spectral basis functions, with 55 corresponding to the Blue Photometer (BP) and 55 to the Red Photometer (RP) \citep{gaia_bprp_processing}. The coefficient form of the spectra can itself in some sense be considered a form of data compression and feature extraction; although the \textit{Gaia} team provides software to convert the coefficients into mean spectra (\texttt{GaiaXPY}\footnote{\url{https://gaia-dpci.github.io/GaiaXPy-website/}}), these converted spectra contain no more information than the coefficients themselves (i.e., because they are converted from the coefficients and their errors). For this reason, we use the coefficients as the input to our neural network, as has most previous work to extract stellar parameters and abundances from the BP/RP spectra. The values of the BP and RP coefficients vary with apparent magnitude, which we do not want to effect the predictions of our network. Thus, we scale each BP and RP coefficient to the first BP or RP coefficient, respectively, associated with that star, as the first coefficients contain the information regarding the overall shape and scaling of the spectrum:
$$\mathrm{BP_{i,scaled}}=\mathrm{BP_{i,original}}/\mathrm{BP_{1,original}}$$
and likewise for the RP coefficients. After this scaling, the values of the first BP and RP coefficients are 1 for all stars, so we remove them from the array of BP/RP coefficients. Thus, the input predictors for our neural network are a set of 108, rather than the original 110, coefficients. \citet{fallows_sanders_bprp} employ a similar procedure to normalize the BP/RP coefficients; however, unlike their work, we do not include photometric colors (ex. from 2MASS or WISE) in the vector of our predictors.

We further normalize the coefficients using the $1$st and $99$th percentiles of the coefficients, i.e.:
$$\mathrm{BP_{i,final}}=\frac{\mathrm{BP_{i}-BP_{i,1\%}}}{\mathrm{BP_{i,1\%}-BP_{i,99\%}}}$$
wherein $\mathrm{BP}_{i,1\%}$ is the $1$st percentile and $\mathrm{BP}_{i,99\%}$ is the $99$th percentile of all of the $i$th BP coefficients in the training data. We perform the same procedure for the RP coefficients.

\begin{figure*}
	\includegraphics[width=6.7in, alt={In the illustrative example of BP/RP spectra, the high-[N/O] spectrum has much deeper absorption features in the CN bands than the [N/O]-typical spectrum, whereas the carbon-rich stellar spectrum has stronger absorption still in both the CN, CH, and C2 features. The NH band and Al lines are not obviously visible in the spectra.}]{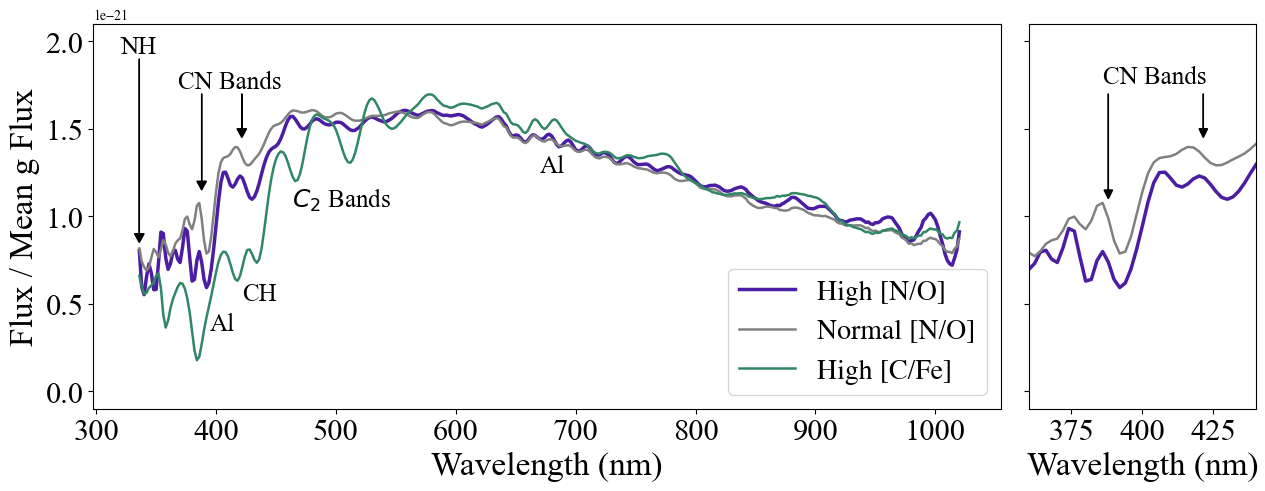}
    \caption{Sample \textit{Gaia} BP/RP spectra, with [N/O]-normal ($\textrm{[N/O]=-0.066}$, grey), high-[N/O] ($\textrm{[N/O]=1.098}$, purple) and, [N/O] normal, high-[C/Fe] ($\textrm{[C/Fe]=1.3}$, dark green) stars with comparable $T_{\textrm{eff}}$ and [Fe/H]. The \textit{Gaia} IDs are 1454785210768079744, 2259240659843862016, and 2982933097213087616, respectively, with $T_\mathrm{eff}\approx5000$~K, $\log g\approx2.4$, and $\textrm{[Fe/H]}\approx-1.35$. The left panel shows the whole BP/RP spectra as converted from the basis coefficients with \textit{GaiXPy}, and the right panel shows the magnified region of the high- and normal-[N/O] spectra containing the CN bands. The abundances of the carbon-rich star are from \citet{Yoon_cemp_2016}, while the other two were selected from APOGEE abundances. Arrows indicate the carbon and nitrogen molecular features, and the strongest aluminum lines in the BP/RP wavelength range are also marked.}
    \label{fig:example_bprp}
\end{figure*}

In summary, the training data for our neural network are comprised of a prediction features vector of the 108 standardized BP and RP coefficients (with the first coefficients removed after scaling) and labels in the form of a five-dimensional vector of APOGEE measurements of $T_{\textrm{eff}}$, $\log g$, [Fe/H], [N/O], and [Al/Fe] for the star corresponding the BP/RP coefficients in the features.

We believe that it is possible to gauge nitrogen abundances from the BP/RP spectra due to the presence of the two CN bands within the wavelength range of the \textit{Gaia} Blue Photometer and also potentially via the NH band near the lower wavelength limit of the Blue Photometer. The recent work of \citet{fallows_sanders_bprp} to estimate nitrogen abundances from the BP/RP spectra also supports this assertion.
To illustrate the nitrogen information in the BP/RP spectra, we provide sample spectra in Fig.~\ref{fig:example_bprp} converted from the coefficients with \texttt{GaiaXPY}. The stars share similar stellar parameters but differ in their [N/O] abundances, with a high-[N/O] star in purple and a nitrogen-typical star in gray. The CN and NH bands are indicated with arrows and are magnified in the right panel of Fig.~\ref{fig:example_bprp} and are visibly more prominent in the spectrum of the nitrogen-rich star as compared to the nitrogen-typical star. We further include the spectrum of a carbon-enhanced star to show that high-[N/O] and high-[C/Fe] stars can be distinguished from their BP/RP spectra using the CH and C$_2$ bands, which is especially important given that we believe that most of the nitrogen information in the BP/RP spectra comes from the CN bands. The flux in the blue wavelengths of the spectrum of the carbon-rich star is noticeably lower than that of the two carbon-typical stars, making the distinction between nitrogen- and carbon-enhanced stars possible by eye from their BP/RP spectra. As we discuss in more detail in Section~\ref{sec:validation}, we do not find carbon enhancement to be a confounding factor in our selection of high-[N/O] candidates.

The results from \citet{ting_2018} suggest that these same CN, CH, and $\textrm{C}_2$ molecular features in the BP/RP spectra may contain the oxygen information which is part of our [N/O] predictions. The oxygen information in the carbon molecular features arises from the fact that most of the oxygen in stellar atmospheres exists in CO; thus, the oxygen abundance directly influences the abundance of other carbon-bearing molecules. \citet{ting_2017} leveraged the oxygen information inherent in the carbon features to determine carbon abundances from optical spectra without oxygen lines, although notably the $R\approx1800$ LAMOST spectra they used are nonetheless much higher resolution than the BP/RP spectra. \citet{ting_2018} note that the effect of oxygen on the molecular carbon features is non-degenerate with the carbon and nitrogen abundance, which is particularly important given that we seek [N/O] predictions as a classifier of 2P stars. We tested predicting [N/Fe] and [O/Fe] as separate output features rather than in a combined [N/O] prediction and found a negligible difference in performance.

Regarding aluminum, which is especially difficult to measure from stellar spectra and does not have immediately obvious absorption lines in the BP/RP spectra (see Fig.~\ref{fig:example_bprp}), we discuss in Section~\ref{sec:additional_tests} that the [Al/Fe] predictions of our network may in fact be closely linked to Na abundances. \citet{Hattori_2024} produces $\alpha$-element predictions from BP/RP spectra which they find arise from information near the Na D absorption lines at 589 and 589.6~nm. They further indicate that some $\alpha$-element information arises from the Mg I line at 516~nm, which is consistent with our findings that our [Al/Fe] predictions are distinct from [$\alpha$/Fe] values (discussed in Section~\ref{sec:additional_tests}). As Na is also enhanced in 2P GC stars \citep[for example, see][]{milone_multiple_pops_GCs, Carretta_2009a, Carretta_2009b}, Na information in our [Al/Fe] predictions would nonetheless be useful for identifying new 2P candidates. We cannot discount the possibility that our [Al/Fe] predictions may in part arise from a series of complicated correlations with other abundances and spectral features; we address this possibility throughout the paper.

One final feature that could affect our identification of GC stars is the helium enhancement present in some 2P stars. He abundance differences between cluster populations correlate strongly with GC mass, with some larger clusters exhibiting significant He enrichment among their second generation members \citep{milone_multiple_pops_GCs}. He abundances can have a substantial effect on a stellar spectrum, with a He-enhanced star being far bluer than a comparable He-typical counterpart \citep[see Figure 4 of][]{Milone_2018}. It is thus possible that He enhancement may impact our network's identification of some 2P-type stars, with the different spectral energy distribution being a readily identifiable feature. He enhancement may also affect $T_\textrm{eff}$ predictions for such stars.

\subsection{Data-Driven Identification of High-[N/O] Stars}
\label{sec:algorithm}

Our machine learning model is a multi-layer perceptron (MLP) --- a type of neural network --- which we parameterise using PyTorch \citep{pytorch}.
Our MLP has an input layer with 108 nodes, each corresponding to one of the BP/RP coefficients, two hidden layers of 128 nodes each with rectified linear unit (ReLU) activation functions, followed by an output layer.
Each hidden layer has a dropout fraction of 0.2 initialized, wherein $20\%$ of the nodes are randomly not used in each forward pass.
Dropout \citep{Hinton_dropout} is a common technique used to prevent overfitting to the training data for neural networks, and our dropout rate of 0.2 was selected after testing rates within the range 0.05 to 0.2 and examining the training and validation loss curves during network training to prevent overfitting.
Per \citet{Gal_2015_dropout}, we use dropout during evaluation as well as training and make 100 predictions per BP/RP spectrum. These individual predictions, which can be thought of as samples from the distribution of predictions, are then averaged for the final inferred abundances and variances used to classify 2P-type candidate stars. This technique helps address uncertainty arising from the initialization of the MLP itself and also has been shown to improve network performance.
\citet{fallows_sanders_bprp} employ the same technique and provide a thorough discussion of model variance in their work.
We use the \texttt{Adam} optimizer \citep{adam} with a learning rate of 0.001. All other \texttt{Adam} hyperparameters are set to the PyTorch defaults.
We train with a batch size of 128.
The learning rate and batch size were set following testing to minimize the loss on the validation set (with additional checks to ensure the false negative and false positive rates were not excessively high, see Section~\ref{sec:validation}).

Heteroscedastic regression is a form of regression that takes into account non-uniform variance among the data \citep[Section 5.3.4 of][]{Simon_Prince_textbook}; we elect to use this form of regression given that we naturally expect stellar parameters and abundances to have non-uniform variance (ex. for abundance measurements to have different variances based on $T_{\textrm{eff}}$, metallicity, etc.). To perform heteroscedastic regression, we use a loss proportional to the negative log-likelihood of a parameterised Gaussian, wherein for each batch in the training the loss is:
$$\text{loss} = \frac{(\mu - y_{\text{true}})^2 }{\sigma^2} + \log \sigma^2$$
In the loss function above, $\mu$ is the vector of the network’s predicted values for each of the five parameters and abundances used as labels, $\log \sigma^2$ is the corresponding logarithm of the variances on each of those predictions, and $y_{\text{true}}$ is the true value of each parameter or abundance. Note that the variances are inferred by the network and not given as part of the training labels. From here forward, we refer to the network’s mean prediction of a parameter or abundance by $\mu_{\text{value}}$ and correspondingly to the network’s predicted variance on the value as $\sigma_{\text{value}}^2$. For instance, we refer to the predicted [N/O] abundance as $\mu_{\textrm{[N/O]}}$ and the corresponding variance as $\sigma_{\textrm{[N/O]}}^2$.The loss function is such that when the residual is large, it dominates the loss; when the residual is low or the variance is particularly high, the $\log \sigma^2$ term dominates the loss. The consideration of the variance in the data within the loss function helps to decrease bias in predictions (i.e. a "regression towards the mean" wherein low values are over-predicted and high values are under-predicted). Moreover, we use the network's predicted variances to help determine which $\mu_{\text{value}}$ predictions are reliable (see Section~\ref{sec:validation}). \citet{fallows_sanders_bprp} stellar parameter predictions from the BP/RP spectra utilized a comparable loss function.

We make note of the fact that a machine learning model's estimations of abundances and stellar parameters are different from actually ``measuring'' these values, as the models are sufficiently complex to learn complicated correlations between various features (i.e., a general correlation of N with metallicity or $\alpha$-element abundance, among many other features). This capability of machine learning approaches to learn correlations between features in the data may itself be a strength in the context of real relationships between abundances (i.e. the relationship between O and carbon molecular features discussed above), particularly given the low resolution of the BP/RP spectra with limited visible atomic and molecular features.

\subsection{Red Giant Sample}
\label{sec:andrae_giants}

Because our model is trained solely on red giants from APOGEE, we likewise can only produce predictions for giants. \citet{andrae_metallicities} produce a reliable selection of RGB stars (see Table 2 of the referenced paper) to which we apply our neural network to make predictions of [N/O] and [Al/Fe] as well as new predictions of $\log g$, $T_{\textrm{eff}}$, and [Fe/H]. The \citet{andrae_metallicities} catalog also includes predictions for $\log g$, $T_{\textrm{eff}}$, and [Fe/H], but we include our own new predictions to maintain consistency among our data. This consistency is especially relevant given that \citet{andrae_metallicities} uses a very different machine learning model \citep[\texttt{XGBoost},][]{xgboost} as compared to our neural network. 

For the MLP to provide reasonable predictions, it is essential that the data to which we apply the network is well represented in our training data. For that reason, we apply the same cuts from the \textit{Gaia} catalog that were applied to our training and validation data in Section~\ref{sec:data} and remove any stars that were in our training or validation data. Note, however, that \citet{andrae_metallicities} make the cut $\texttt{parallax}/\texttt{parallax\_error}>4$, which is slightly more restrictive than the cut we use in our training and validation data. We apply an additional cut on the $\log g$ prediction produced by \citet{andrae_metallicities}, $\log g_{\textrm{XGBOOST}}<3.0$, again with the goal of maintaining consistency with our training data from APOGEE. \citet{andrae_metallicities} already apply a cut of $T_{\textrm{eff,XGBOOST}}<5200$~K, which is consistent with the cut we made on $T_{\textrm{eff,APOGEE}}$ in the training and validation data. Our ability to rely on the \texttt{XGBOOST} predictions of $T_{\textrm{eff}}$ and $\log g$ to ensure consistency with our training data motivates our use of the vetted \citet{andrae_metallicities} RGB catalog for our selection of candidates. For additional cuts applied to the vetted RGB catalog, the reader is referred to \citet{andrae_metallicities}. Finally, we apply an extra cut to remove spectra with high extinction with $\texttt{ebv}<1$ \citep[from][]{sfd_ebv}; although this cut is not included in the training data to maximize the number of samples we train with, stars with $\texttt{ebv}<1$ are nonetheless well represented within the training and validation data. By contrast, high-extinction sources are relatively less well-represented in the training data, and to ensure high quality predictions, we thus remove these stars. Perhaps even more importantly, high extinction significantly impacts sensitivity in blue wavelengths, with $\textrm{E}(\textrm{B}-\textrm{V}) = 1.0$ corresponding to a decrease of about 3.5 magnitudes around 430~nm, near the CN features. For this reason, we construct a high-reliability sample with $\texttt{ebv}<0.2$ in addition to the full catalog. With the exclusion of stars in our APOGEE training and validation data, we are left with 6\,878\,665 RGB stars in the full catalog to which we apply our MLP.

\section{Network Performance}
\label{sec:results}

In this section, we describe several tests that validate the results of our neural network. These tests include the performance of our network on the 20\% of the APOGEE-BP/RP data that we withheld from training and performance checks on the selection of new candidates from the \citet{andrae_metallicities} RGB catalog and from a catalog of giants from GALAH \citep{galah_dr3, galah}.

\subsection{Validation Data Performance}
\label{sec:validation}

\begin{figure*}
	\includegraphics[width=6.2in, alt={The predictions of the neural network for each of the stellar parameters in the validation set approximately follow the 1-to-1 line with their corresponding values from APOGEE, although there is a bias to over-predict low values of [N/O] and [Al/Fe] and to under-predict high values. The bias is worse for the aluminum predictions.}]{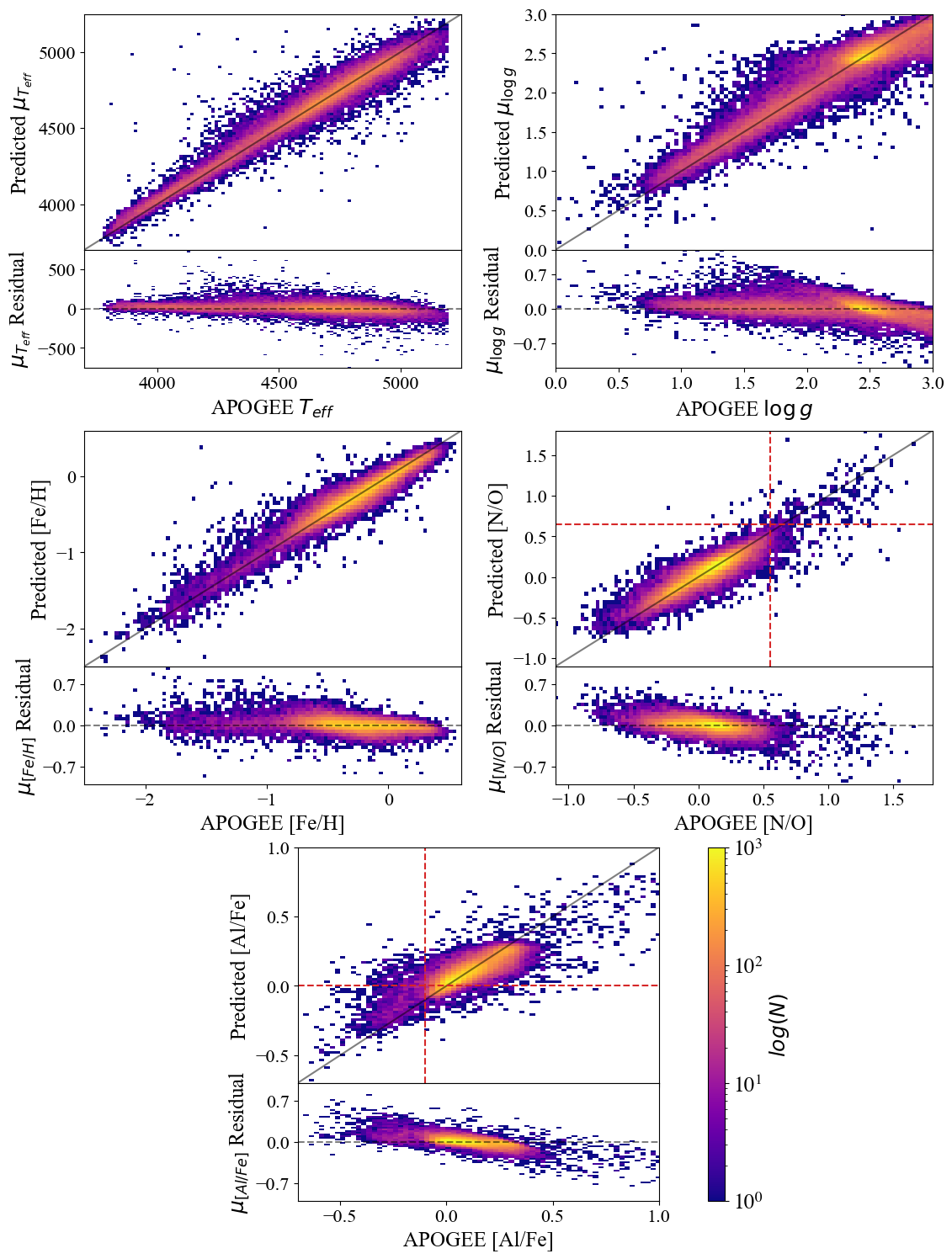}
    \caption{Moving from left to right and top down, the validation performance of the network predictions for $\mu_{T_{\textrm{eff}}}$, $\mu_{\log g}$, $\mu_{\textrm{[Fe/H]}}$, $\mu_{\textrm{[N/O]}}$, and $\mu_{\textrm{[Al/Fe]}}$. The large upper panel of each subplot depicts the prediction versus the true value from APOGEE with the 1:1 line marked in black. The narrow lower panel is the prediction residual (i.e., $\mu_{value}-value_{APOGEE}$) with the horizontal black line marking a residual of zero. The colorbar is shared for all panels and marks the color-mapping of the histograms as the log-scaled number of stars per pixel. In the panels depicting the [N/O] and [Al/Fe] predictions, the vertical lines depict the true cutoff values we use from APOGEE abundances to classify stars as high-[N/O] at $\textrm{[N/O]}=0.55$ and $\textrm{[Al/Fe]}=-0.1$, respectively. The horizontal red dashed lines represent the cutoffs we use in our predicted values of $\mu_{\textrm{[N/O]}}$ and $\mu_{\textrm{[Al/Fe]}}$ at 0.65 and 0.0, respectively, as is described in Section~\ref{sec:validation}; note that these lines do not account for $\sigma_{\textrm{[N/O]}}$ and $\sigma_{\textrm{[Al/Fe]}}$, which are also used in our selection cuts.}
    \label{fig:validation_performance}
\end{figure*}

Our test data is comprised of 39\,816 stars with both BP/RP spectra and APOGEE abundances. We show the validation performance of our network in Fig.~\ref{fig:validation_performance}, separated by each of the five stellar parameters or abundances we predict. As is evident in the first three panels of Fig.~\ref{fig:validation_performance}, the predictions of $\log g$, $T_{\textrm{eff}}$, and [Fe/H] appear to well follow the 1:1 line with the true APOGEE values; this result is consistent with many previous works which find these fundamental stellar parameters to be well-predicted from the BP/RP coefficients through various data-driven approaches \citep[e.g.][among others]{andrae_metallicities,fallows_sanders_bprp}. We note a small population of stars with over-predicted surface gravities at $\log g\approx2$, which could be related to the red clump. There is also a slight systematic over-prediction of metallicities at [Fe/H]~$\lesssim-1$, with an average residual of $+0.07$ dex for stars with [Fe/H]~$\lesssim-1$. The root mean squared error (RMSE) of $\mu_{T_{\textrm{eff}}}$, $\mu_{\log g}$, and $\mu_{\textrm{[Fe/H]}}$ are 62.4~K, 0.162, and 0.096, respectively.

The [N/O] and [Al/Fe] predictions also approximately follow the 1:1 lines with the APOGEE values well. However, examination of the residuals of these predictions in the bottom panels in Fig.~\ref{fig:validation_performance} do reveal a bias towards the median (i.e. over-prediction of low values and under-prediction of high values). This behavior is consistent with the trends in [N/Fe] prediction discussed in \citet{fallows_sanders_bprp}. This trend is a natural behavior of many machine learning models and is unsurprising given the unbalanced nature of our training data wherein we have relatively few high-[N/O] stars. However, we argue that this bias is not exceedingly problematic for our use case given that we use the $\mu_{\textrm{[N/O]}}$ and $\mu_{\textrm{[Al/Fe]}}$ predictions only to produce a binary classification of stars as 2P-type or not 2P-type rather than directly as abundances, as we will discuss in more detail shortly. It is worth noting that the bias appears to be the most significant for the $\mu_{\textrm{[Al/Fe]}}$ estimates, which may be related to the relative difficulty of identifying Al (or Na) information in the BP/RP spectra as compared to N or O (as discussed in Section~\ref{sec:data}). The RMSE of $\mu_{\textrm{[N/O]}}$ is 0.096, and the RMSE of $\mu_{\textrm{[Al/Fe]}}$ is 0.076.

In Fig.~\ref{fig:no_stdev}, we examine the MLP's output standard deviation of the [N/O] prediction, $\sigma_{\textrm{[N/O]}}$, in the validation data as a function of $T_{\textrm{eff}}$, $\log g$, [Fe/H], [N/O], and the $\mu_{\textrm{[N/O]}}$ residual. There are several trends in the value of $\sigma_{\textrm{[N/O]}}$ worth noting. First, $\sigma_{\textrm{[N/O]}}$ increases noticeably at higher temperatures ($T_{\textrm{eff}}\gtrsim4700$~K. This trend is likely physically motivated given that the CN bands grow weaker in the spectra of hotter stars. Second, $\sigma_{\textrm{[N/O]}}$ is also higher at [Fe/H]~$\lesssim-1$. We believe this behavior may be caused both by the fact that there are fewer metal-poor than metal-rich stars in our training data and also because the CN bands become less prominent in stars at lower metallicities \citep[e.g., as recently shown in][]{carretta_2024_discrepancies}. $\sigma_{\textrm{[N/O]}}$ is consistently high for stars with [Fe/H]~$\approx-2$. Interestingly, the trend of $\sigma_{\textrm{[N/O]}}$ with the prediction residual in $\mu_{\textrm{[N/O]}}$ is somewhat less distinct than the trends with temperature and metallicity, though stars with higher magnitudes of prediction residuals do tend to have slightly higher values of predicted $\sigma_{\textrm{[N/O]}}$ as well. This trend coincides with the higher values of $\sigma_{\textrm{[N/O]}}$ in stars with true [N/O] abundances that are well above or below the median value of [N/O] in our training and validation data. The behavior is typical of machine learning algorithms given that these values of [N/O] are comparably less represented in our training data as compared to stars with [N/O]~$\approx0.1$, which drives both the residuals and uncertainties for these less common stars to be higher. Finally, there is a useful linear correlation between $\sigma_{\textrm{[N/O]}}$ and the reported [N/O] error from APOGEE, which is marked by the black line and which we utilize to make our selection of high-[N/O] candidates (as described in Section~\ref{sec:candidate_selection}).

Inferred standard deviations of the $T_{\textrm{eff}}$, $\log g$, [Fe/H], and [Al/Fe] are included in Appendix~\ref{appendix}.

\begin{figure}
	\includegraphics[width=\columnwidth, alt={The network [N/O] standard deviation increases at effective temperatures greater than about 5000 K, metallicities less than -1, and at the lowest and highest APOGEE [N/O] values. The standard deviation also correlates with the APOGEE [N/O] error and slightly less strongly with the residual on the [N/O] prediction.}]{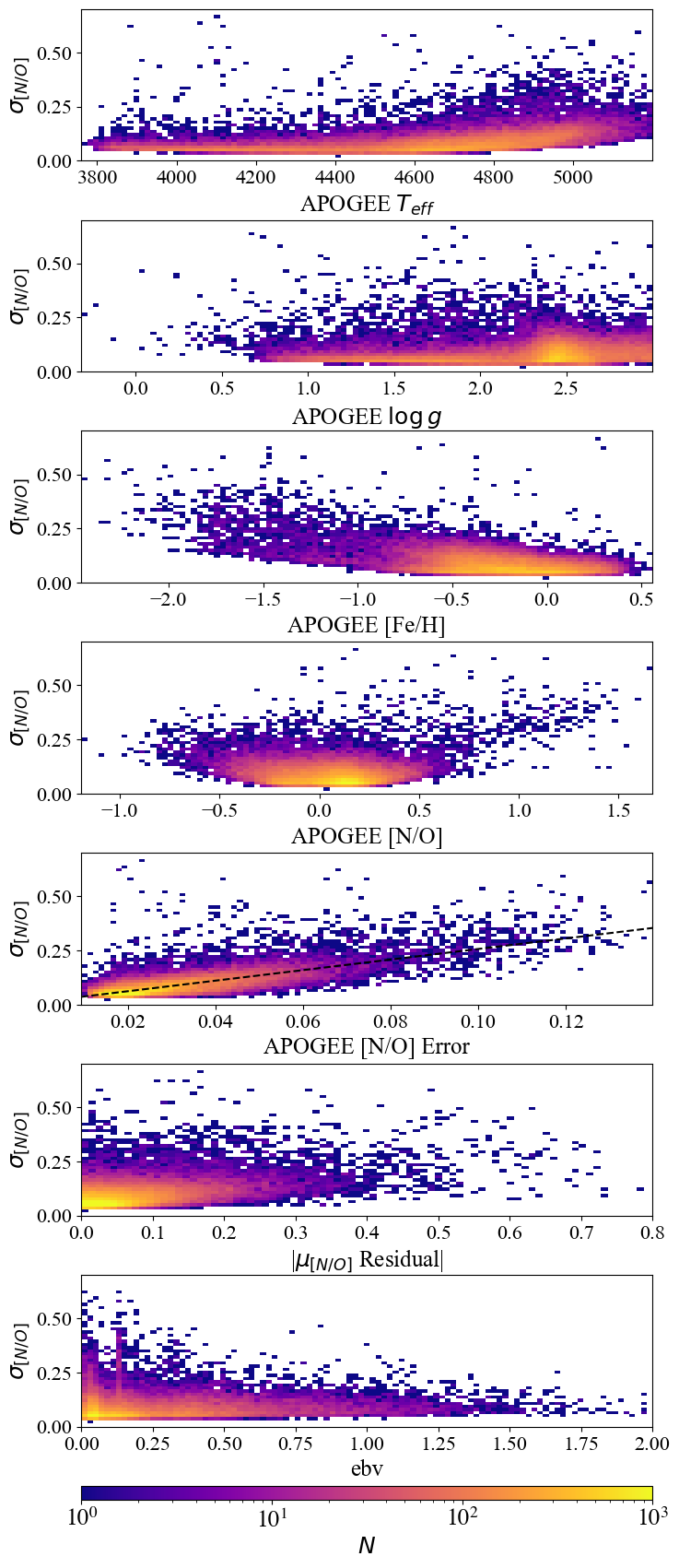}
    \caption{From top down, the MLP standard deviation prediction of [N/O] ($\sigma_{\textrm{[N/O]}}$) versus APOGEE values for $T_{\textrm{eff}}$, $\log g$, [Fe/H], [N/O], the residual of the $\mu_{\textrm{[N/O]}}$ prediction in the validation dataset, and $\textrm{E}(\textrm{B}-\textrm{V})$ from \citet{sfd_ebv} The colorbar is shared for all panels and marks the color-mapping of the histograms as the log-scaled number of stars per pixel. The black line in the fourth row depicts the best-fit line relating the APOGEE [N/O] error to $\sigma_{\textrm{[N/O]}}$, $\textrm{APOGEE [N/O] error}=0.19\times\sigma_{\textrm{[N/O]}}+0.013$.}
    \label{fig:no_stdev}
\end{figure}

\subsection{High-[N/O] Candidate Selection}
\label{sec:candidate_selection}
We use the following criteria to select high-[N/O] candidate stars using our network predictions:
\begin{itemize}
  \item $\mu_{\textrm{[N/O]}}-0.19\times\sigma_{\textrm{[N/O]}}>0.65$
  \item $\mu_{\textrm{[Al/Fe]}}-\sigma_{\textrm{[Al/Fe]}}>0.$
  \item $\mu_{T_{\textrm{eff}}}<5000$
  \item $\mu_{\textrm{[Fe/H]}}>-2.0$
\end{itemize}
The cut on predicted temperature has two motivations, one physical and one empirical. First, it is known that the CN bands grow weak or disappear altogether at $T_\mathrm{eff}\gtrsim5000$~K, and since we believe that most of the nitrogen information in the BP/RP spectra comes from those CN bands, there is good reason to exclude stars with higher temperatures from our sample of high-[N/O] candidates. Second, our network predicts increasing $\sigma_{\textrm{[N/O]}}$ values at high $T_{\textrm{eff}}$ (Fig.~\ref{fig:no_stdev}), suggesting greater uncertainties in the [N/O] predictions at higher temperatures. This result is unsurprising given the behavior of the CN bands at higher temperatures and further justifies the exclusion of warm giants from our high-[N/O] candidates. The cut on metallicity as part of our criteria for high-[N/O] candidates is motivated by the metallicity distribution of the training APOGEE-BP/RP cross-match, which contains very few stars with [Fe/H]$<2.0$. As compared to the rest of the stars in the training data, proportionally more of these very metal-poor stars were high-[N/O] stars \citep[as is also noted in][]{belokurov_kravstov_nitrogen}, biasing the neural network to predict high $\mu_{\textrm{[N/O]}}$ values for very metal-poor stars. We felt for this reason that our [N/O] predictions for stars with [Fe/H]$<-2.0$ may be unreliable and thus exclude them from our sample of high-[N/O] candidates.

The [N/O] selection at $\mu_{\textrm{[N/O]}}-0.19\times\sigma_{\textrm{[N/O]}}>0.65$ takes advantage of the observed correlation between $\sigma_{\textrm{[N/O]}}$ and the APOGEE [N/O] error in Fig.~\ref{fig:no_stdev}. By approximating the relation between $\sigma_{\textrm{[N/O]}}$ and the APOGEE [N/O] error as linear, we take the slope of the best-fit line (shown in Fig.~\ref{fig:no_stdev}), 0.19, as a factor for  $\sigma_{\textrm{[N/O]}}$ in our cut; thus, $\mu_{\textrm{[N/O]}}-0.19\times\sigma_{\textrm{[N/O]}}$ is intended to be approximately analogous to APOGEE [N/O]$-$[N/O] error. One high outlier in $\sigma_{\textrm{[N/O]}}$ was excluded in the calculation of the relation between $\sigma_{\textrm{[N/O]}}$ and the APOGEE [N/O] error. We note that the thresholds in $\mu_{\textrm{[N/O]}}$ and $\mu_{\textrm{[Al/Fe]}}$ that we use to select high-[N/O] candidates are slightly higher than the cuts we use to make the true classification from APOGEE abundances (as described in \ref{sec:data}). These higher thresholds are motivated by the fact that our [N/O] and [Al/Fe] predictions, although robust, do sometimes have errors of a few tenths of a dex, and our goal is to construct a pure, reliable sample of new high-[N/O] candidates. Thus, to maintain a low false positive rate in our selection of new candidates, we use more stringent cuts on $\mu_{\textrm{[N/O]}}$ and $\mu_{\textrm{[Al/Fe]}}$ to select stars with true abundances of $\textrm{[N/O]}>0.55$ and $\textrm{[Al/Fe]}>-0.1$. These cuts were selected after testing with our validation data to maintain a balance of a low contamination rate without losing most or all of the true 2P stars from our classification.

\begin{figure}
	\includegraphics[width=\columnwidth, alt={In the confusion matrix for the validation data, there were 39461 true negatives, 7 false positives, 252 false negatives, and 96 true positive detections of high-[N/O] stars.}]{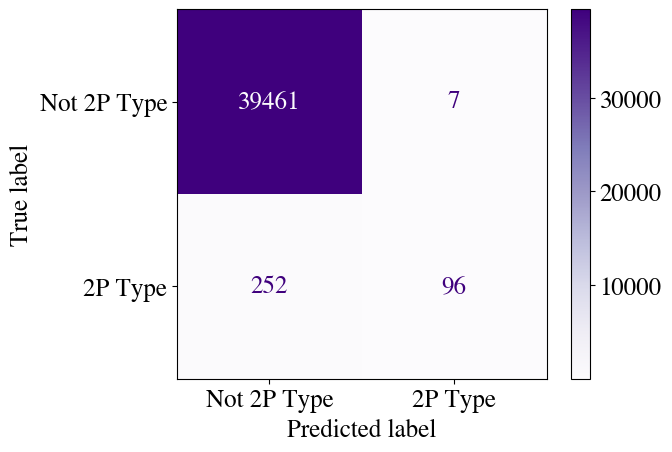}
    \caption{The confusion matrix of classifications of stars in the validation dataset, with true labels on the vertical axis generated from APOGEE abundances (as described in Section~\ref{sec:data}) and the predicted labels made from the network predictions as described in Section~\ref{sec:candidate_selection}. The 2P-type label refers to high-[N/O], high-[Al/Fe] stars consistent with having formed in the second generation of GCs, while the not 2P-type label refers to all other stars.}
    \label{fig:confusion_matrix}
\end{figure}

Applying these criteria to our validation data, we classify stars as either "2P-type" or "not 2P-type" with over 99\% accuracy. The confusion matrix showing the results of our classification in the validation data is presented in Fig.~\ref{fig:confusion_matrix}. 103 stars are classified as having 2P chemistry with a 6.8\% false positive rate. We note, however, that due to the rarity of 2P stars in the Galaxy and thus within our validation data, this false positive rate corresponds to only 7 out of 39\,468 non-2P-type stars being misclassified as stars with 2P [N/O] and [Al/Fe] abundances. Inspection of these 7 false positive classifications in the validation data reveal that 3 have APOGEE $\textrm{[N/O]}>0.55$ but have errors sufficiently high such that they do not satisfy the condition $\textrm{[N/O]}-\textrm{[N/O] error}>0.55$. The remaining 4 false positives all have [N/O] abundances close to the threshold, with all having reported APOGEE $\textrm{[N/O]}>0.49$. It appears that these stars are very nitrogen-rich but not sufficiently oxygen-depleted to satisfy the [N/O] requirement. All false positives clear the [Al/Fe] threshold at -0.1. We note that carbon overabundance does not appear to be a source of false positives, as the mean [C/Fe] of the false positive classifications in the validation data is -0.16. This result is reassuring given that we believe the main source of nitrogen information in the BP/RP spectra comes from the CN bands, with the NH band barely being in the range of the \textit{Gaia} Blue Photometer.

Notably, our classification of 2P-type stars in the validation data has a somewhat substantial false negative rate, with 72\% of true 2P-type stars being misclassified as not 2P-type. Some of these misclassifications arise not only from $\mu_{\textrm{[N/O]}}$ and $\mu_{\textrm{[Al/Fe]}}$ predictions inconsistent with our selection criteria but also from the exclusion of warm or very metal-poor giants from our classification of 2P-type stars; naturally, some true 2P-type stars will be hotter than 5000~K or more metal-poor than $\textrm{[Fe/H]}=-2.0$. Our false negative rate is approximately consistent with that of the identification of carbon-enhanced metal-poor (CEMP) stars from the BP/RP spectra in \citet{lucey_cemp}. The false negative rate is a necessary side effect of maintaining high purity in our candidate selection, and given the rarity of known 2P-type stars in the Galactic field, even with this false negative rate our sample of candidates is multiple times larger than the comparable sample that can be compiled from APOGEE data (see Section~\ref{sec:results}). Thus, since we nonetheless construct a large sample of candidates, we find having a high purity in our selection to be a worthwhile trade-off for the false negative rate. However, we will publish our full catalog of predictions, enabling the use of less stringent candidate selections by the community if desired. The catalog of our predictions is detailed in Appendix~\ref{catalog}.

\subsection{Additional Tests of High-[N/O] Candidates}
\label{sec:additional_tests}

In addition to testing our neural network on validation data from APOGEE, we perform several additional tests to confirm the reliability of the sample of our new high-[N/O] candidates. First, we identify stars from \citet{andrae_metallicities} that are in GCs according to \citet{vasiliev_gc} and compare the high-[N/O] fraction among these cluster members to the full set of field giants. We find that 9.82\% of GC members are classified by our algorithm as high-[N/O] stars as compared to just 0.16\% of field giants in the halo--a difference of well over a magnitude. This result is consistent with the expectation that GCs should contain many more 2P stars \citep[ex.][]{Milone_2017}, which are very rare in the field. These fractions are likely to be artificially low due to the relatively high false positive rate in our classification; approximately 72\% of true 2P stars in our validation data were misclassified as not high-[N/O]. For comparison, if we make classifications for 2P-type stars with APOGEE abundances in our combined training and validation data, we find that 31.5\% of giants in GCs are classified as 2P-type as compared to just 0.43\% in the halo.

\begin{figure}
	\includegraphics[width=\columnwidth, alt={All but one of the high-[N/O] candidates are sodium-enhanced (the outlier has [Na/Fe] slightly below 0) and most are oxygen depleted compared to the overall population of metal-poor stars. The majority of high-[N/O] candidates are identified as GC members.}]{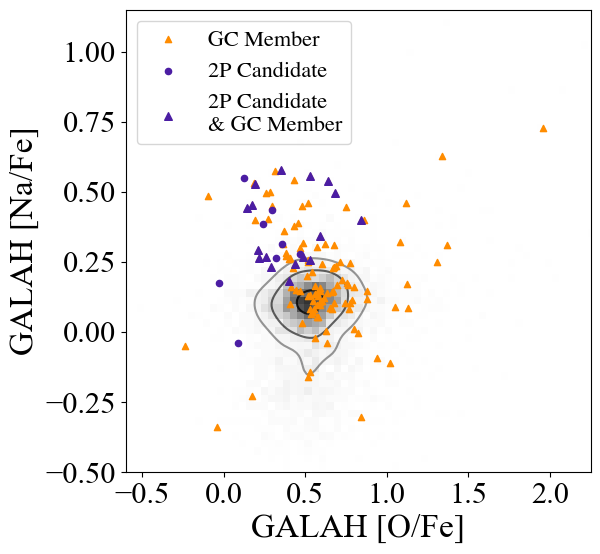}
    \caption{The figure shows the sodium versus oxygen abundances of stars in our GALAH dataset. The background gray histogram represents the overall distribution of [Na/Fe]-[O/Fe] abundances among GALAH stars with $\textrm{[Fe/H]}<-0.5$, with metal-poor stars selected for a more realistic comparison with GCs, which tend to be metal-poor. The grey contours marking 30\%, 50\%, and 90\% contours. The small orange triangles mark members of GCs in the GALAH data \citep{vasiliev_gc}. The purple circles represent stars that we identify as high-[N/O] candidates, and the large purple triangles represent stars that are both high-[N/O] candidates and GC members.}
    \label{fig:galah_gc}
\end{figure}

\begin{figure*}
	\includegraphics[width=6.6in, alt={The is clearly no correlation between GALAH [Al/Fe] and the network's inferred [Al/Fe]. Interestingly, the network's [Al/Fe] abundances split into two sequences relative to GALAH metallicity (a low and high aluminum sequence). Of the other elements, only [Na/Fe] clearly correlates with the inferred [Al/Fe] abundances, with a cut at [Al/Fe]=-0.1 splitting the stars into low- and high-sodium groups.}]{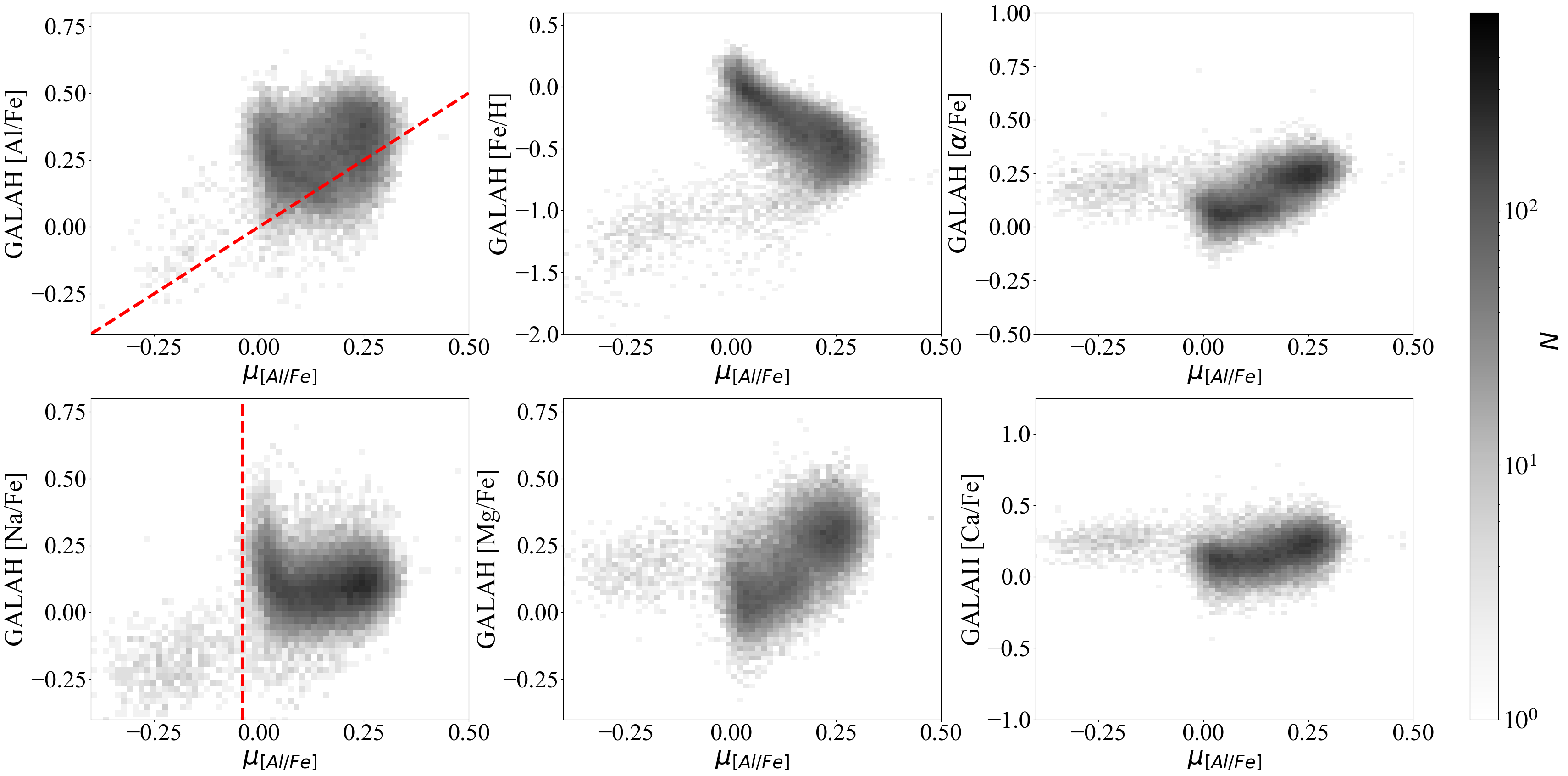}
    \caption{Moving from left to right and beginning with the top row, GALAH [Al/Fe], [Fe/H], [$\alpha$/Fe], [Na/Fe], [Mg/Fe], and [Ca/Fe] versus $\mu_{\textrm{[Al/Fe]}}$. The GALAH [Al/Fe] versus $\mu_{\textrm{[Al/Fe]}}$ has the 1:1 line depicted in red, and the GALAH [Na/Fe] versus $\mu_{\textrm{[Al/Fe]}}$ has a vertical red line at $\mu_{\textrm{[Al/Fe]}}=-0.04$ to depict the cutoff between low- and high-Na stars in the abundance plane. Panels are log-scaled in density and share a colorbar.}
    \label{fig:galah_abundances}
\end{figure*}

Our second test of the predictive power and accuracy of our network utilizes the GALactic Archaeology with HERMES \citep[GALAH,][]{galah_dr3, galah, galah_data_reduction, galah_data_reducation_2} survey, which provides abundances for several light elements associated with the GC abundance anti-correlations, including Na, O, Al, Mg, and C. To further validate our candidate selection, we select stars from GALAH that have been identified as GC members in \citet{vasiliev_gc} and explore their light element abundances. We apply several cuts to the data, beginning by selecting giants by using the cuts $T_\mathrm{eff}<5200$~K and $\log g<3$. Stars with potentially erroneous stellar parameters are removed via the cut $\texttt{flag\_sp}=0$, problematic metallicity measurements removed with $\texttt{flag\_fe\_h}=0$, and high signal-to-noise observations are selected with $\texttt{snr\_c3\_iraf}>30$. When other element abundances are used, we also apply a cut for $\texttt{flag\_X\_h}=0$. Each of these criteria is used at the recommendation of the GALAH collaboration's best practices\footnote{\url{https://www.galah-survey.org/dr3/using_the_data/}}. We also apply the same cuts from the \textit{Gaia} that we describe in Section~\ref{sec:data}, including a cut for $\texttt{ebv}<0.2$ for consistency with our high-reliability sample of candidates, and by necessity, our sample is limited to stars with BP/RP spectra available from \textit{Gaia} DR3. Stars that are present in our training or validation data are removed. After these cuts, we have 30\,604 stars with both GALAH data and BP/RP spectra.

In Fig.~\ref{fig:galah_gc}, we show the [Na/Fe]-[O/Fe] abundance patterns of stars in our GALAH sample. The distribution of the overall population of giants is depicted in the 2D histogram in the background with GC member giants \citep{vasiliev_gc} over-plotted on top. The light element anti-correlation of GC members is evident, with a group of GC member (1G) stars falling within the regime of typical GALAH stars. Another group of GC member stars (2P) is clearly outside the limits of the typical GALAH stars in the [Na/Fe]-[O/Fe] plane, with evident Na-enhancement and O-depletion relative to GC 1G stars, as is expected for GC 2P stars \citep[e.g., as described in the review in][]{milone_multiple_pops_GCs}. We then use the BP/RP spectra of all stars in this sample to identify high-[N/O] candidates from our GALAH sample using the same procedure outlined in Section~\ref{sec:validation}. We note that our candidates generally populate the Na-enhanced, O-depleted region of the abundance space, again as one would expect of second generation GC stars. The fraction of GC members in GALAH identified as high-[N/O] candidates, 12.8\%, is much higher than the overall fraction of high-[N/O] candidates that we identify in GALAH stars that are not GC members in \citet{vasiliev_gc}, just 0.046\%. As in our test of GC members above, this result is consistent with the fact that GCs ought to contain more high-[N/O] and high-[Al/Fe] stars than the MW field. We recognize that there are many 2P GC stars that our algorithm does not tag as high-[N/O], which is consistent with our recovery of high-[N/O] stars having a false negative rate of approximately 72\%. When we relax the cuts for candidate selection for this sample to $\mu_{\textrm{[N/O]}}>0.55$ and $\mu_{\textrm{[Al/Fe]}}>-0.1$, we find that we recover most of the apparent 2P stars from this figure as well as many more field giants in that region of the [Na/Fe]-[O/Fe] plane. However, with those relaxed selection cuts we also tag a small but noticeable number of field stars outside the Na-enriched regime of the abundance space which we assume are false positives, which is consistent with our tests in Section~\ref{fig:validation_performance}. We find our original cuts at $\mu_{\textrm{[N/O]}}>0.65$ and $\mu_{\textrm{[Al/Fe]}}>0.1$ to thus be well-suited to this test, too, although this result may suggest that many more 2P stars can be found in the field from our predicted abundances if one relaxes the selection cuts from those we use here.

\begin{figure}
	\includegraphics[width=\columnwidth, alt={In the [Fe/H>-0.5 group, there is no clear correlation between GALAH [Na/Fe] and and the inferred aluminum abundances, but in the [Fe/H]<-0.5, there is a weak to moderate correlation between the two.}]{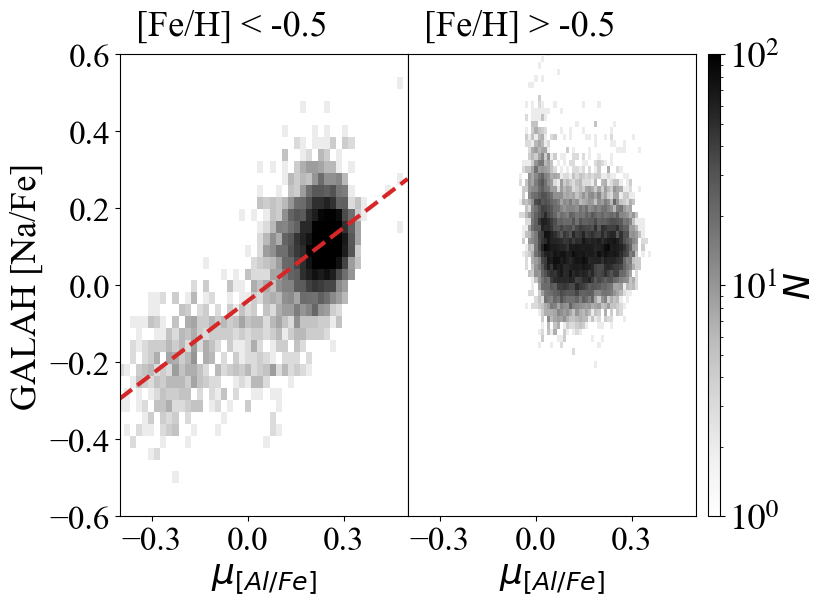}
    \caption{GALAH [Na/Fe] versus $\mu_{\textrm{[Al/Fe]}}$ for GALAH [Fe/H]~$<-0.5$ (left) and[Fe/H]~$>-0.5$ (right). Both panels are log-scaled in density and share a colorbar. The red dashed line in the left panel is the best-fit line relating $\mu_{\textrm{[Al/Fe]}}$ and GALAH [Na/Fe].}
    \label{fig:galah_aluminum}
\end{figure}

Finally, Al abundances in the optical rely on only three lines, one of which (669.8~nm; see Fig.~\ref{fig:example_bprp}) can appear quite weak in metal-poor stars. Thus, because aluminum abundances are notoriously difficult to derive from spectra and because \textit{Gaia} BP/RP spectra are so low resolution, the $\mu_{\textrm{[Al/Fe]}}$ predictions demand further examination. We again use GALAH abundances to further explore the nature of the $\mu_{\textrm{[Al/Fe]}}$ predictions; in particular, we are looking at whether the aluminum predictions of our neural network are correlated with other similar elements, in spite of the relatively good agreement between our $\mu_{\textrm{[Al/Fe]}}$ and the APOGEE [Al/Fe] measurements in Fig.~\ref{fig:validation_performance}. In Fig~\ref{fig:galah_abundances}, we examine $\mu_{\textrm{[Al/Fe]}}$ as a function of [Al/Fe], [Fe/H], [$\alpha$/Fe], [Na/Fe], [Mg/Fe], and [Ca/Fe] from GALAH. The correlation between $\mu_{\textrm{[Al/Fe]}}$ clearly does not follow the 1:1 line with the GALAH [Al/Fe] measurement as well as is seen in our validation data, though this behavior can be explained given that GALAH aluminum abundances are known to differ from those reported by APOGEE \citep[see][]{Buder_GALAH}.
As is apparent in the top right panel of Fig~\ref{fig:galah_abundances}, there is a correlation between $\mu_{\textrm{[Al/Fe]}}$ and [$\alpha$/Fe], and this is again likely to be expected as aluminum is predominantly produced in core-collapse supernovae \citep{Kobayashi_origin_of_elements} and thus can be expected to generally correlate with $\alpha$-element abundances. Likewise, there is some relation between metallicity and our $\mu_{\textrm{[Al/Fe]}}$ prediction, but the nonmonotonicity of that relation suggests that our neural network is not purely correlating one with the other. Interestingly, there are clearly multiple tracks in $\mu_{\textrm{[Al/Fe]}}$ as a function of metallicity.

The correlation of $\mu_{\textrm{[Al/Fe]}}$ with [Mg/Fe] and [Ca/Fe] is present but less distinct, which is again perhaps a result of the fact that all of these elements share a nucleosynthetic source in Type II supernovae. However, in the [Na/Fe]-$\mu_{\textrm{[Al/Fe]}}$ plane (bottom left panel of Fig~\ref{fig:galah_abundances}), there is a distinct and valuable pattern. By placing a cut at $\mu_{\textrm{[Al/Fe]}}=0$, as we illustrate with the vertical red line, stars are very clearly divided into low- and high-Na groups. This clean separation is not possible with any of the other elements that we examine with GALAH and suggests that there may in fact be information on [Na/Fe] in the $\mu_{\textrm{[Al/Fe]}}$ values. We further examine the relation between $\mu_{\textrm{[Al/Fe]}}$ and [Na/Fe] for metal-poor stars ([Fe/H]~$<-0.5$) and metal-rich stars ([Fe/H]~$>-0.5$) in Fig.~\ref{fig:galah_aluminum}. We find that $\mu_{\textrm{[Al/Fe]}}$ particularly well-predicts Na abundance for metal-poor stars, though the relationship is less strong for metal-rich stars; given that most 2P-type stars, including our candidates, tend to be metal-poor, this is a positive result. Moreover, if the cut on $\mu_\textrm{[Al/Fe]}$ is removed for candidates, three more candidates, seemingly false positives, appear in Fig.~\ref{fig:galah_gc}; two have $\textrm{[Na/Fe]}<0$, further suggesting that there is Na information in the $\mu_\textrm{[Al/Fe]}$ predictions. It has been previously observed that Al and Na abundances trace each other in GC light element anticorrelations \citep[e.g.][among many others]{Carretta2010}, indicating that this may be a useful and sensible result.

In short, we believe that there is likely information on several elements, especially $\alpha$-elements, in our $\mu_{\textrm{[Al/Fe]}}$ predictions. In particular, we find that a cut at $\mu_{\textrm{[Al/Fe]}}=0$ cleanly separates stars into low- and high-[Na/Fe] populations in data from GALAH. The correlation between $\mu_{\textrm{[Al/Fe]}}$ and GALAH [Na/Fe] is especially evident at [Fe/H]$<-0.5$. Given that most of our high-[N/O] stars exist at lower metallicities and that Na-enhancement is also present in 2P stars, it is perhaps unsurprising that our cut for high $\mu_{\textrm{[Al/Fe]}}$ values helps to select 2P stars. In validation testing, including the cut of $\mu_{\textrm{[Al/Fe]}}-\sigma_{\textrm{[Al/Fe]}}>0.$ removes an additional 6 false positive stars, and with numbers of high-[N/O] stars so low, this removal of 6 false positives is relevant to our contamination rate. Moreover, as we discuss in Section~\ref{sec:spatial_kinematic}, these $\mu_{\textrm{[Al/Fe]}}$ predictions even produces a sensible, if imperfect, separation between \textit{in-situ} and accreted stars.

\section{Properties of Galactic High-[N/O] Candidates}
\label{sec:nrich_properties}

In this section, we examine the properties of our high-[N/O] candidates. We are particularly interested in the distribution and kinematics of our high-[N/O] candidates within the Galactic halo, which allows us to explore the contribution of GCs to the early Galaxy \citep[][]{belokurov_aurora, rix_poor_old_heart, conroy_galactic_disk_h3}. To remove thin disk stars, which are more likely to be N-enhanced from sources other than an origin in GCs \citep{belokurov_kravstov_nitrogen}, we convert \textit{Gaia} astrometric data into spherical Galactocentric positions and velocities, and we apply a cut at $v_\phi<160$~km/s. Radial velocities are from \textit{Gaia}, and the distances are calculated from parallaxes. We use \texttt{Astropy}'s \citep{astropy:2013,astropy:2018,astropy:2022} default \texttt{Galactocentric} reference frame, which has a Solar distance of 8.122~kpc from the Galactic center and a Solar velocity of (12.9, 245.6, 7.78)~km/s in Cartesian Galactocentric coordinates. After applying the selection for halo stars and a cut at $\texttt{ebv}<0.2$ in addition to the selection criteria for high-[N/O] candidates that we outline in Section~\ref{sec:validation}, we identify \ncandidates~new high-[N/O], high-[Al/Fe] candidates in the Galactic halo that are not associated with a GC in \citet{vasiliev_gc}. This low extinction selection constitutes our high-reliability sample and is shown in Fig.~\ref{fig:sky_map}.  If all \nrgb~RGB stars with $\texttt{ebv}<1$ from \citep{andrae_metallicities} are used, we identify 1\,432 new high-[N/O], high-[Al/Fe] candidates in the Galactic halo. We publish this full catalog but encourage caution regarding predictions for highly extincted stars due to the decreased sensitivity at blue wavelengths, and for the remainder of this paper we use our high-reliability, low extinction sample. By contrast to our \ncandidates~new candidates in the low extinction sample, our selection of APOGEE giants has only 133 high-[N/O] stars with $v_\phi<160$~km/s that are not associated with a GC (using the cuts $\textrm{[N/O]}-\textrm{[N/O] error}>0.55$ and [Al/Fe]~$>-0.1$, as specified in Section~\ref{sec:data}). Thus, our candidate sample is over six times larger than the existing comparable catalog of known high-[N/O] stars from APOGEE.

\begin{figure}
	\includegraphics[width=\columnwidth, alt={In the 2D histogram representing the halo as a whole in our sample, the cut-out removal of the Galactic disk on the sky is evident. The high-[N/O] candidates are spread across the sky, with the highest concentrations present near the Galactic center.}]{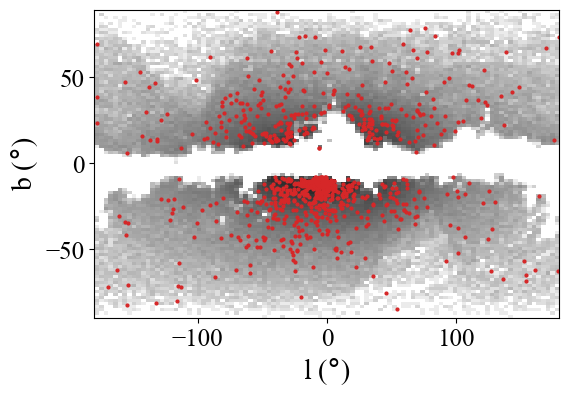}
    \caption{In red points, the distribution of high-[N/O] and high-[Al/Fe] candidates in Galactic latitude and longitude. The grey 2D histogram is the log-scaled distribution of all stars not in GCs that pass the cut at $v_{\phi}<160~\textrm{km/s}$. Note that as a result of this cut and the extinction cut, low Galactic latitutdes are excluded.}
    \label{fig:sky_map}
\end{figure}

\begin{figure*}
	\includegraphics[width=6.6in, alt={In both the energy-angular momentum plane and the histogram of energies, both the high aluminum and the high-[N/O] candidates clearly occupy lower energies than the low aluminum stars. Almost all of the high aluminum stars and most of the high-[N/O] candidates lie below (at lower energies) the in situ and accreted line of separation in the energy-angular momentum plane, whereas the low aluminum stars span the plane more broadly and are not confined to the accreted area of the plane. The distribution of energies of the high aluminum and nitrogen-rich candidates are very similar.}]{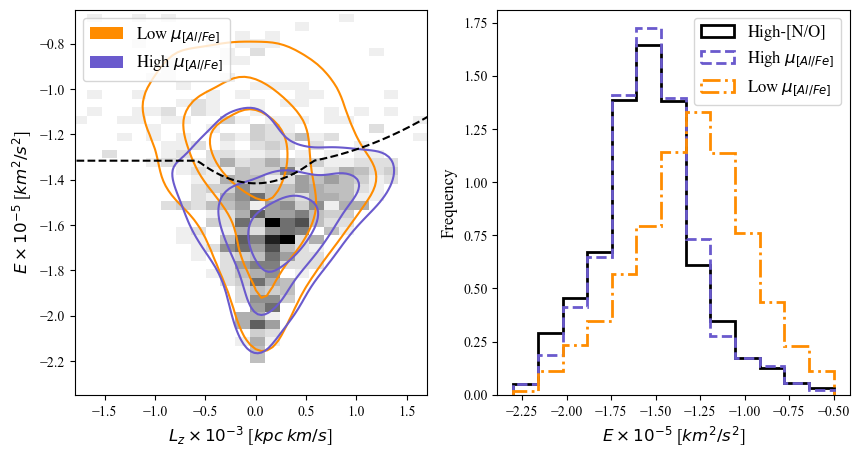}
    \caption{Left: The separation of low-$\mu_{\textrm{[Al/Fe]}}$ accreted and high-$\mu_{\textrm{[Al/Fe]}}$ \textit{in-situ} (\textit{Aurora}) field stars from our sample in energy-$L_{z}$ space. The orange and blue contours depict the 20\%, 50\%, and 80\% contours of the low-$\mu_{\textrm{[Al/Fe]}}$ and high-$\mu_{\textrm{[Al/Fe]}}$, and the black dashed line marks the kinematic separation adopted from \citet{belokurov_kravstov_nitrogen}. The grey 2D histogram is the distribution of high-[N/O] candidates in the E-$L_{z}$ space. Metallicities are restricted to $-1.4<\mu_{\textrm{[Fe/H]}}<-1.1$, where separation of accreted and \textit{in-situ} stars based on [Al/Fe] is most effective. Right: The frequency of high-[N/O] candidates (black solid line), high-$\mu_{\textrm{[Al/Fe]}}$ (blue dashed line), and low-$\mu_{\textrm{[Al/Fe]}}$ (orange dash-dotted line) stars in the halo as a function of orbital energy. The same [Fe/H] restrictions are used.}
    \label{fig:insitu_separation}
\end{figure*}

\begin{figure}
	\includegraphics[width=\columnwidth, alt={The fraction of high-[N/O] candidates relative to the total halo increases from 2 times 10^-3  at 10 kpc to about 1 times10^-2 at 1 kpc from the Galactic center. The radial trend of the fraction of high-[N/O] candidates relative to accreted stars has almost the same shape but is about an order of magnitude higher. By contrast, the fraction of high-[N/O] candidates relative to the in-situ halo is much flatter, only increasing from about 2 times 10^-1 at 10 kpc to 4 times 10^-1 at 1 kpc from the Galactic center.}]{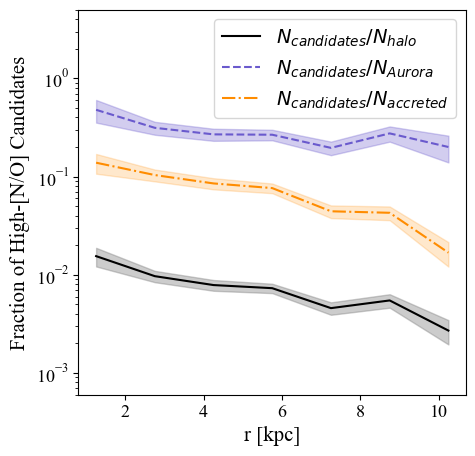}
    \caption{Ratio of high-[N/O] candidates to total number of stars in the halo (black line), \textit{in-situ} or \textit{Aurora} stars (dashed blue line), and accreted stars (dashed orange line) as a function of distance from the Galactic center in spherical Galactocentric coordinates. The shaded regions indicate the statistical uncertainties. We maintain our selection for $-1.4<\mu_{\textrm{[Fe/H]}}<-1.1$.}
    \label{fig:radial_fraction}
\end{figure}

\subsection{Spatial and Kinematic Distribution of High-[N/O] Candidates}
\label{sec:spatial_kinematic}

In the left panel of Fig.~\ref{fig:insitu_separation}, we explore the distribution of our high-[N/O] candidates in the total orbital energy and vertical component of the angular momentum, $E-L_{z}$ space. Energies are calculated using the \texttt{MilkyWayPotential} from \texttt{gala} \citep{gala,adrian_price_whelan_2020_4159870,galpy}. [Al/Fe] abundances have previously reliably been used to separate stars that formed \textit{in situ} in the Galaxy from those that were accreted with other structures \citep{hawkins_aluminum}, with the MW's satellites having consistently lower aluminum abundances than \textit{in-situ} stars \citep{Hasselquist_satellite_abundances}. With this in mind, we attempt to use our $\mu_{\textrm{[Al/Fe]}}$ predictions to separate accreted and \textit{in-situ} stars. We impose rather strict cuts on $\mu_{\textrm{[Al/Fe]}}$ to attempt a clean selection, motivated by the higher bias in the $\mu_{\textrm{[Al/Fe]}}$ prediction noted in Section~\ref{sec:validation} and the notable but imperfect trend with [Na/Fe] observed in Fig.~\ref{fig:galah_aluminum}. Thus, low-aluminum "accreted" stars are selected via $\mu_{\textrm{[Al/Fe]}}<-0.2$, and high-aluminum "\textit{in-situ} stars" are selected to have $\mu_{\textrm{[Al/Fe]}}>0.2$. We also impose a cut on metallicity, $-1.4<\mu_{\textrm{[Fe/H]}}<-1.1$, as this is the metallicity range in which accreted and \textit{in-situ} stars can be best separated by aluminum abundances \citep{belokurov_kravstov_nitrogen}. We maintain the cut at $\mu_{T_\mathrm{eff}}<5000$ for both samples and make the same velocity cuts that we applied to our high-[N/O] candidates to remove thin disk stars. The black dashed line marks the empirical separation of accreted and \textit{in-situ} (\textit{Aurora}) stars developed in \citet{belokurov_kravstov_nitrogen}, which is:
\begin{align*}
E=-1.316\:&:\:L_{z}<-0.58 \\
E=-1.416+0.3L_{z}^2\:&:\:-0.58<L_{z}<0.58 \\
E=-1.341+0.075L_{z}^2\:&:\:L_{z}<0.58
\end{align*}
with $L_z$ in units of  $10^{-3}~\mathrm{kpc~km/s}$ and $E$ in units of $10^{-5}~\mathrm{km^2/s^2}$. The energy values are shifted by $-0.016\times10^{-5}~\mathrm{km^2/s^2}$ to account for the different energies of \texttt{gala}'s \texttt{MilkyWayPotential} and the Milky Way model used in \citet{belokurov_kravstov_nitrogen}. This separation of accreted and \textit{in-situ} stars was developed using the observed differences in APOGEE [Al/Fe] abundances between accreted and \textit{in-situ} stars. Interestingly, other chemical elements also show strong differentiation across this $E-L_z$ divide. For example, \citet{Monty2024} demonstrate that both field stars and MW Globular Clusters display distinct levels of [Si/Fe], [Eu/Fe] and [Eu/Si] on either side of the boundary. We primarily use this orbital separation as a means with which to judge the efficacy of our identification of accreted and \textit{in-situ} stars using $\mu_{\textrm{[Al/Fe]}}$. 

Notably, our selection of stars with high $\mu_{\textrm{[Al/Fe]}}$ predictions appears to be a fairly clean sample of \textit{in-situ} stars, as the contours of this sample lie almost entirely within the kinematic selection in the $E-L_z$ space. This behavior is especially remarkable given the aforementioned difficulty of measuring aluminum spectroscopically and reinforces the suggested [Al/Fe] and [Na/Fe] information in the $\mu_{\textrm{[Al/Fe]}}$, both of which are useful to distinguish accreted and \textit{in-situ} stars \citep{hawkins_aluminum, Nissen_Schuster_2010, Das_2020}. By contrast, the selection of low $\mu_{\textrm{[Al/Fe]}}$ "accreted" stars is notably more contaminated than the high $\mu_{\textrm{[Al/Fe]}}$ selection, although the contours do fall above the line of separation more than the \textit{in-situ} stars. We conclude that the $\mu_{\textrm{[Al/Fe]}}$ values contain information of abundances that are useful to separate accreted and \textit{in-situ} stars, albeit imperfectly. It is especially reassuring that the \textit{in-situ} selection appears to be pure, as this result suggests that the stars identified as high-[Al/Fe] by our network are actually high-[Al/Fe], as is typical of \textit{in-situ} MW stars. Given that we use a selection for high-[Al/Fe] to identify 2P candidates, this result can be taken as yet another form of validation of our selection. The apparent contamination of the "low-[Al/Fe], accreted" stars is less concerning in this context as it will not affect the purity of our selection of high-[Al/Fe] 2P candidates; the fact that high-[Al/Fe] stars are sometimes tagged as low-[Al/Fe] may be related to the false negative rate discussed in Section~\ref{sec:candidate_selection}.

Finally, the distribution of high-[N/O] candidates, shown by the grey 2D histogram, clearly falls mostly among the \textit{in-situ} regime of the $E-L_z$ space defined both by the kinematic separation and the high $\mu_{\textrm{[Al/Fe]}}$ contours. This result is consistent with the previous findings of \citet{belokurov_kravstov_nitrogen} that most high-[N/O] field stars from APOGEE are members of \textit{Aurora}. The connection between the high-[N/O] candidates and selected \textit{in-situ} stars can be further observed in the right panel of Fig.~\ref{fig:insitu_separation}, which shows the energy distributions of the high-[N/O], \textit{in-situ}, and accreted candidates. There is a clear difference in the energy distribution of the \textit{in-situ} and accreted candidates, with the high $\mu_{\textrm{[Al/Fe]}}$ stars tending to have lower energies than the low $\mu_{\textrm{[Al/Fe]}}$ stars, which is consistent with the expected behavior from \textit{in-situ} and accreted stars, respectively. Notably, the high-[N/O] candidate energy distribution closely traces that of the \textit{in-situ} candidates, with many of the candidates having lower orbital energies than the accreted stars. This trend is again indicative of the \textit{in-situ} and high-[N/O] populations being linked and is also in accordance with results from \citet{belokurov_kravstov_nitrogen}. We perform a 2-sample Kolmogorov–Smirnov (K-S) test to compare the energy distributions of the high-[N/O] candidates with the \textit{in-situ} and accreted distributions and find that the high-[N/O] and \textit{in-situ} energy distributions are approximately consistent with having been drawn from the same populations, producing a p-value of 0.455 and a test statistic of 0.037. By contrast, the K-S test results of the high-[N/O] and accreted energy distributions indicate that these two groups differ significantly, producing a p-value of $1.5\times10^{-101}$ and a test statistic of 0.380. The results of the K-S tests further suggest a close association between the high-[N/O] candidates and \textit{in-situ} stars.

\begin{figure*}
	\includegraphics[width=6.3in, alt={The fraction of N-rich stars among both APOGEE and the BP/RP candidates is approximately flat (with a slight decrease) at 2 times 10^-2 at metallicities beginning at [Fe/H]=-2, but the fraction drops rapidly to just above 10^-4 at about [Fe/H]=-1. There is a slight increase again at metallicities greater than -0.5, with the increase being higher for the APOGEE stars than the BP/RP candidates. The distribution of high-[N/O] candidates is noticeably more metal-poor than the distribution of the halo as a whole.}]{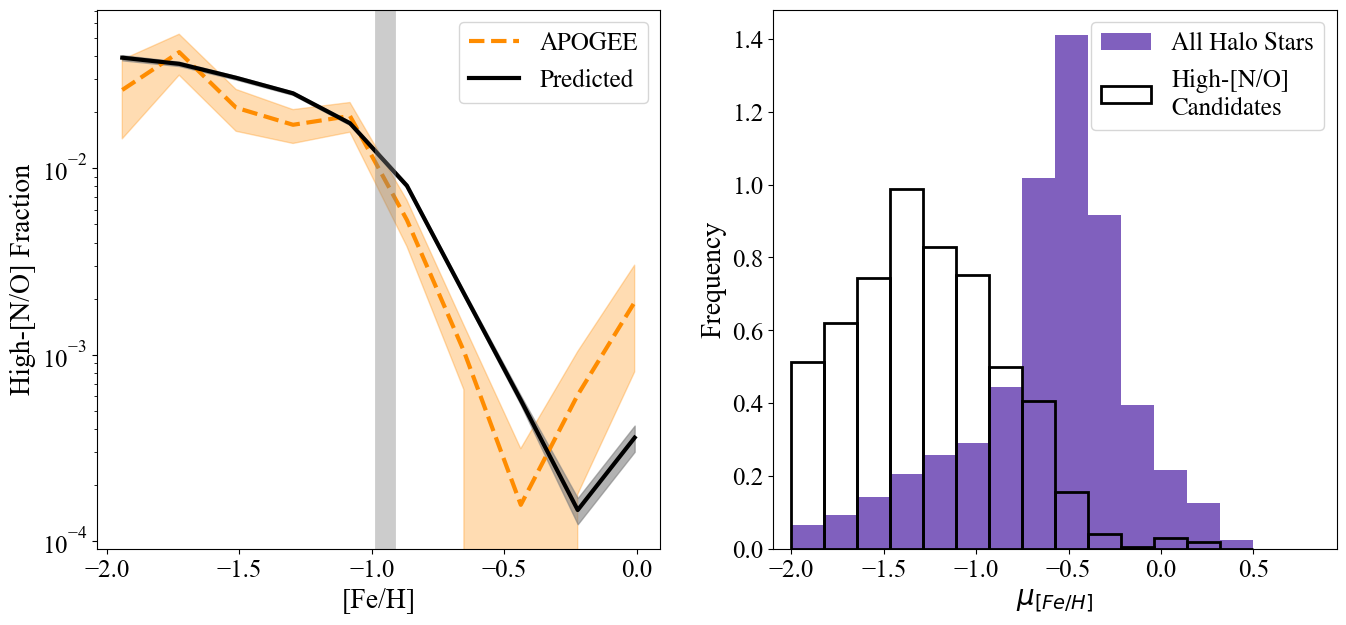}
    \caption{Left: The fraction of high-[N/O] field stars in the halo as a function of metallicity. The orange dashed line shows the data from APOGEE using APOGEE metallicities, and the black line is constructed from our sample of candidates with $\mu_{\textrm{[Fe/H]}}$ as the metallicity. The high-[N/O] fraction from our candidates is multiplied by a factor of 3.6, which is used to represent our projected false positive rate of 72\% from the validation data. The shaded regions surrounding the trend lines indicate the statistical uncertainties; note that the uncertainties from our candidate data set are much smaller due to the larger sample size compared to APOGEE. The completion of spin-up, the formation of the Galactic disk, is marked by the vertical gray shading at $-1.0\lesssim$~[Fe/H]$\lesssim-0.9$ \citep{belokurov_aurora}. Right: The normalized metallicity distribution using $\mu_{\textrm{[Fe/H]}}$ of all giants in our sample (purple) and the candidate high-[N/O] stars (black outline).}
    \label{fig:metallicity}
\end{figure*}

We examine the radial distribution of the high-[N/O] candidates within the Galaxy in Fig.~\ref{fig:radial_fraction}.  We see the expected increase in the fraction of high-[N/O] stars with decreasing Galactocentric radius in the halo as a whole as noted in \citet{Horta_nrich_stars, Schiavon_2017, Martell_2011, belokurov_kravstov_nitrogen}. Our false negative rate results in systematically lower fractions of high-[N/O] stars as compared to those quoted in \citet{Horta_nrich_stars}; regardless, the general trend recovered is similar. The central concentration of the 2P-type candidates is also evident in the sky distribution in Fig.~\ref{fig:sky_map}.

Fig.~\ref{fig:radial_fraction} also shows the ratio of high-[N/O] candidates to accreted and \textit{in-situ} stars, which were classified according to the chemical criteria described above. We notice an increasing trend in the fraction of high-[N/O] candidates to accreted stars, with the fraction growing by a factor of $\approx10$ from $r\approx10$~kpc to $r\approx1.25$~kpc. The large increase in this fraction at small Galactocentric radii is in agreement with the results from \citet{belokurov_kravstov_nitrogen}, although they found that the fraction increased by a factor of $\approx40$. We believe that the rate of the increase in the fraction of high-[N/O] stars relative to the accreted candidates approaching the Galactic center is more similar to the fraction relative to the halo overall due to the considerable contamination in our selection of accreted stars, as is evident in the right panel of Fig.~\ref{fig:insitu_separation}. This contamination makes our sample of accreted candidates more similar to the halo as a whole rather than being a distinct subgroup, as is the case with the accreted stars selected via APOGEE abundances in \citet{belokurov_kravstov_nitrogen}. As a result, the trend of 2P-type candidates to accreted halo stars does closely resemble that of the ratio relative to the halo overall. We recover a flatter trend in the ratio of high-[N/O] candidates to \textit{in-situ} stars, as noted by \citet{belokurov_kravstov_nitrogen}. The more similar radial distribution of 2P-type candidates and \textit{in-situ} halo stars further indicates that these candidates are predominately members of \textit{Aurora}. As we noted in Fig.~\ref{fig:insitu_separation}, the \textit{in-situ} selection with $\mu_{\textrm{[Al/Fe]}}$ appears mostly pure, suggesting that this result is particularly robust.

With regards to all radial distributions, our selection criteria for 2P-type, \textit{in-situ}, and accreted stars differ from those used by \citet{belokurov_kravstov_nitrogen} by necessity to maintain relatively reliable selections. For these reasons, the fractions of high-[N/O] stars that we recover relative to the halo, \textit{it-situ} halo, and accreted halo cannot be directly compared to the results of that work, although the trends are nonetheless very similar.

\subsection{Metallicity Distribution of the High-[N/O] Candidates}
\label{sec:metallicity}

We now discuss the metallicities ($\mu_{\textrm{[Fe/H]}}$) of our high-[N/O] field giants. We show the metallicity distribution of our candidates in the right panel of Fig.~\ref{fig:metallicity} and the corresponding high-[N/O] fraction as a function of metallicity in the left panel. The high-[N/O] candidates are comprised of a larger fraction of metal-poor stars as compared to the overall population of RGB stars. As a consequence, the high-[N/O] fraction drops with increasing metallicity, while remaining at a low rate by [Fe/H]~$\approx-0.8$. For comparison, we also show the high-[N/O] fraction as a function of metallicity using APOGEE abundances for stars in our training and validation data.  We note that using {\it Gaia} BP/RP spectrophotometry we recover a trend extremely similar to that in \citet{belokurov_kravstov_nitrogen}, who also use APOGEE data. "Spin-up," the formation of the MW disk \citep{belokurov_aurora}, is marked with the horizontal gray line in Fig.~\ref{fig:metallicity}. By the metallicity of spin-up ([Fe/H]~$\approx-0.9$, the high-[N/O] fraction is much lower than it was at lower metallicities.

We echo the interpretation of \citet{belokurov_kravstov_nitrogen} regarding these results. Namely, the production of high-[N/O] stars, which currently consensus suggests is a yet-unknown process taking place between 1G and 2P stars in GCs, was much more prevalent in star formation at lower metallicities. Thus, at lower metallicities and hence further in the MW's past, GCs contributed a much higher fraction of stellar production in the Galaxy. Around the time of spin-up, that contribution, traced by the high-[N/O] fraction, dropped precipitously.

With regards to this discussion of metallicity in particular, we note that our substantial false negative rate may impact the interpretation of these results. Examining the high-[N/O] candidate recovery rate in both our validation data and in individual GCs at varying metallicities, we find little conclusive evidence that the false negative rate depends strongly on stellar metallicity, except perhaps some indication that the false negative rate may increase slightly at higher metallicities. This behavior of the false negative rate may be responsible for the offset between APOGEE's and our candidate high-[N/O] fraction at [Fe/H]~$\approx0$ in the left panel of Fig.~\ref{fig:metallicity}.  In all likelihood, however, our false negatives are determined by a number of factors including metallicity, effective temperature, the level of [N/O] enhancement, and also features of the BP/RP spectra themselves. To some extent, Fig.~\ref{fig:metallicity} is intended as a scientifically motivated validation of our candidates rather than a new result in itself, and the general agreement between APOGEE and our BP/RP candidates is promising.

\section{Conclusions}
\label{sec:conclusions}

We use a multi-layer perceptron neural network to perform heteroscedastic regression on the \textit{Gaia} BP/RP spectra. Our MLP takes only the BP/RP coefficients as input and predicts stellar $\log g$, $T_{\textrm{eff}}$, [Fe/H], [N/O], and [Al/Fe]. We use the $T_{\textrm{eff}}$, $\log g$, [N/O], and [Al/Fe] predictions to classify candidates with chemistry typical of GC second generation stars (e.g. nitrogen and aluminum overabundance and oxygen depletion, or "high-[N/O] stars" in our phrasing). We show that our predictions of $\mu_{T_{\textrm{eff}}}$, $\mu_{\log g}$, $\mu_{\textrm{[Fe/H]}}$, $\mu_{\textrm{[N/O]}}$, and $\mu_{\textrm{[Al/Fe]}}$ are robust as compared to baseline APOGEE values in our validation data (Fig.~\ref{fig:validation_performance}) and that by selecting stars with $\mu_{\textrm{[N/O]}}-0.19\times\sigma_{\textrm{[N/O]}}>0.65$, $\mu_{\textrm{[Al/Fe]}}-\sigma_{\textrm{[Al/Fe]}}>0.$, $\mu_{T_\mathrm{eff}}<5000$, and $\mu_{\textrm{[Fe/H]}}>-2.0$, we can produce a pure sample of high-[N/O] candidates from their BP/RP spectra. We further validate this selection using GALAH DR3 data by comparing our predicted high-[N/O] fraction in the field to the high-[N/O] fraction in GCs, finding a much higher fraction of high-[N/O] stars among cluster members, as expected.

From our selection, we identify \ncandidates~new field stars in the Galactic halo as high-[N/O] candidates, constituting a sample over six times larger than can be made from APOGEE. We use these new candidates to study the properties of high-[N/O] stars in unprecedented detail. We summarize our findings as follows:
\begin{enumerate}
    \item Our high-[N/O] candidates are present among known GC members at a higher rate than in the Galactic field (Section~\ref{sec:candidate_selection}), and examination of these candidates within the [Na/Fe]-[O/Fe] plane constructed from GALAH abundances shows that they generally fall within the Na-enhanced, O-depleted regime, as is consistent for 2P stars. These results serve to validate our candidate selection.
    \item Exploration of the $\mu_{\textrm{[Al/Fe]}}$ prediction with GALAH abundances reveal that this prediction can be used to separate stars into low- and high-Na groups. Al has previously been found to trace Na in GCs \citep{Carretta2010}, with both being useful to separate stars into accreted and \textit{in-situ} \citep{hawkins_aluminum,Nissen_Schuster_2010, Das_2020}. We thus use $\mu_{\textrm{[Al/Fe]}}$ to make this separation in the $E-L_z$ plane in Fig.~\ref{fig:insitu_separation}, finding that a relatively clean separation is possible between the accreted, low-$\mu_{\textrm{[Al/Fe]}}$ and \textit{in-situ}, high-$\mu_{\textrm{[Al/Fe]}}$ stars. The selection of \textit{in-situ} stars appears to be particularly pure.
    \item Within the $E-L_z$ plane, the majority of the high-[N/O] candidates clearly appear to be associated with the \textit{in-situ} population, which can further be confirmed via examination of their orbital energy and Galactocentric radial distributions (Fig.~\ref{fig:insitu_separation}, \ref{fig:radial_fraction}). We also find the increasing trend in the high-[N/O] fraction in the halo with decreasing Galactocentric radius found by \citet{Horta_nrich_stars, Martell_2011, Schiavon_2017}.
    \item The fraction of high-[N/O] candidates in the halo drops above a metallicity of [Fe/H]~$\approx-1$, which is approximately concurrent with spin-up and is consistent with the findings of \citet{belokurov_kravstov_nitrogen} using APOGEE data.
\end{enumerate}

Our sample of high-[N/O], high-[Al/Fe] candidates represents the current largest collection of stars with chemistries consistent with the second generation of GCs in the Galactic field. Moreover, our approach adds to the mounting community evidence that the \textit{Gaia} BP/RP spectra contain valuable information regarding stellar parameters and can be used to identify chemically peculiar candidates in spite of their low resolution. We intend to leverage the large number of candidates to tag high-[N/O] stars to their GC of origin using their orbital properties, perhaps enabling the discovery of new GC stellar streams. 
 
\section*{Acknowledgements}

We thank the anonymous referee for their helpful review, which has improved the final quality of this paper. SGK thanks Tom Hehir for his useful discussions regarding neural networks. SGK acknowledges PhD funding from the Marshall Scholarship, supported by the UK government and Trinity College, Cambridge.
HZ thanks the Science and Technology Facilities Council (STFC) for a PhD studentship.
AAA acknowledges support from the Herchel Smith Fellowship at the University of Cambridge and a Fitzwilliam College research fellowship supported by the Isaac Newton Trust.

This work made extensive use of the Python packages \texttt{Numpy} \citep{harris2020array}, \texttt{Scipy} \citep{2020SciPy}, \texttt{Matplotlib} \citep{Hunter:2007}, \texttt{Scikit-learn} \citep{scikit-learn}, and \texttt{Gala} \citep{galah,adrian_price_whelan_2020_4159870}. This work made use of \texttt{Astropy}:\footnote{http://www.astropy.org} a community-developed core Python package and an ecosystem of tools and resources for astronomy \citep{astropy:2013, astropy:2018, astropy:2022}.

This work has made use of data from the European Space Agency (ESA) mission
{\it Gaia} (\url{https://www.cosmos.esa.int/gaia}), processed by the {\it Gaia}
Data Processing and Analysis Consortium (DPAC,
\url{https://www.cosmos.esa.int/web/gaia/dpac/consortium}). Funding for the DPAC
has been provided by national institutions, in particular the institutions
participating in the {\it Gaia} Multilateral Agreement. 

This work made use of data from the Apache Point Observatory Galactic Evolution Experiment \citep[APOGEE][]{APOGEE_DR17}. Funding for the Sloan Digital Sky 
Survey IV has been provided by the 
Alfred P. Sloan Foundation, the U.S. 
Department of Energy Office of 
Science, and the Participating 
Institutions. 

SDSS-IV acknowledges support and 
resources from the Center for High 
Performance Computing  at the 
University of Utah. The SDSS 
website is www.sdss4.org.

SDSS-IV is managed by the 
Astrophysical Research Consortium 
for the Participating Institutions 
of the SDSS Collaboration including 
the Brazilian Participation Group, 
the Carnegie Institution for Science, 
Carnegie Mellon University, Center for 
Astrophysics | Harvard \& 
Smithsonian, the Chilean Participation 
Group, the French Participation Group, 
Instituto de Astrof\'isica de 
Canarias, The Johns Hopkins 
University, Kavli Institute for the 
Physics and Mathematics of the 
Universe (IPMU) / University of 
Tokyo, the Korean Participation Group, 
Lawrence Berkeley National Laboratory, 
Leibniz Institut f\"ur Astrophysik 
Potsdam (AIP),  Max-Planck-Institut 
f\"ur Astronomie (MPIA Heidelberg), 
Max-Planck-Institut f\"ur 
Astrophysik (MPA Garching), 
Max-Planck-Institut f\"ur 
Extraterrestrische Physik (MPE), 
National Astronomical Observatories of 
China, New Mexico State University, 
New York University, University of 
Notre Dame, Observat\'ario 
Nacional / MCTI, The Ohio State 
University, Pennsylvania State 
University, Shanghai 
Astronomical Observatory, United 
Kingdom Participation Group, 
Universidad Nacional Aut\'onoma 
de M\'exico, University of Arizona, 
University of Colorado Boulder, 
University of Oxford, University of 
Portsmouth, University of Utah, 
University of Virginia, University 
of Washington, University of 
Wisconsin, Vanderbilt University, 
and Yale University.

This work made use of the Third Data Release of the GALAH Survey \citep{galah_dr3}. The GALAH Survey is based on data acquired through the Australian Astronomical Observatory, under programs: A/2013B/13 (The GALAH pilot survey); A/2014A/25, A/2015A/19, A2017A/18 (The GALAH survey phase 1); A2018A/18 (Open clusters with HERMES); A2019A/1 (Hierarchical star formation in Ori OB1); A2019A/15 (The GALAH survey phase 2); A/2015B/19, A/2016A/22, A/2016B/10, A/2017B/16, A/2018B/15 (The HERMES-TESS program); and A/2015A/3, A/2015B/1, and A/2015B/19, and A/2016A/22, and A/2016B/12, and A/2017A/14 (The HERMES K2-follow-up program). We acknowledge the traditional owners of the land on which the AAT stands, the Gamilaraay people, and pay our respects to elders past and present. This paper includes data that has been provided by AAO Data Central (datacentral.org.au).

This paper made used of the Whole Sky Database (wsdb) created by Sergey Koposov and maintained at the Institute of Astronomy, Cambridge with financial support from the Science $\&$ Technology Facilities Council (STFC) and the European Research Council (ERC).

\section*{Data Availability}

This paper relies on publicly available data from \textit{Gaia} DR3 \citep{gaia_dr3}, APOGEE \citep{APOGEE_DR17}, and GALAH \citep{galah}. The catalog of GC members is publicly available from \citet{vasiliev_gc}. The catalog of our network predictions is available on \href{https://zenodo.org/records/14261030?token=eyJhbGciOiJIUzUxMiJ9.eyJpZCI6IjY0ZmEzZmM3LTFiZDMtNDA0Ni04MGMwLWI4YTYxNDUyMGFiMCIsImRhdGEiOnt9LCJyYW5kb20iOiJiZWI5MDk4NjRkNWYxZjkzZDExMWU5NzMwMzZiNGZhMiJ9.BQtkilNBRSZjDO550ojMlvvSXwI3UHCvs8HqJvkVermx3iiEXIOmoh4okmLumldOiB4owGCKO9cpGkPo4-LNDA}{Zenodo}. The validation data predictions are also provided at the same link for the evaluation of different candidate selection prediction, as desired. A description of the columns in the catalog is included in Appendix~\ref{catalog}.



\bibliographystyle{mnras}
\bibliography{bibliography}




\appendix

\section{Catalog of Predictions}
\label{catalog}

Our catalog of inferred abundances and variances has 12 columns:
\begin{enumerate}
    \item \textit{Gaia} \texttt{source\_id}
    \item \texttt{teff}, \texttt{logg}, \texttt{feh}, \texttt{no}, and \texttt{alfe}, which correspond to $\mu_{T_{\textrm{eff}}}$, $\mu_{\log g}$, $\mu_{\textrm{[Fe/H]}}$, $\mu_{\textrm{[N/O]}}$, and $\mu_{\textrm{[Al/Fe]}}$, respectively
    \item \texttt{teff\_stdev}, \texttt{logg\_stdev}, \texttt{feh\_stdev}, \texttt{no\_stdev}, and \texttt{alfe\_stdev}, which correspond to $\sigma_{T_{\textrm{eff}}}$, $\sigma_{\log g}$, $\sigma_{\textrm{[Fe/H]}}$, $\sigma_{\textrm{[N/O]}}$, and $\sigma_{\textrm{[Al/Fe]}}$, respectively
    \item \texttt{ebv} from \citet{sfd_ebv}, for convenient selection of our high-reliability, low extinction sample
\end{enumerate}

We will also provide as a separate table the variances of the 100 iterations of network predictions for each inferred value (which is described in Sec.~\ref{sec:algorithm} The table of validation predictions is formatted identically with the exception of the \texttt{ebv} column, which is not included.

\section{Additional Inferred Variances}
\label{appendix}

Below we present the values of $\sigma_{T_{\textrm{eff}}}$ (Fig.~\ref{fig:teff_stdev}), $\sigma_{\log g}$ (Fig.~\ref{fig:logg_stdev}), $\sigma_{\textrm{[Fe/H]}}$ (Fig.~\ref{fig:feh_stdev}), and $\sigma_{\textrm{[Al/Fe]}}$ (Fig.~\ref{fig:alfe_stdev}) for our validation data.

\begin{figure}
	\includegraphics[width=\columnwidth, alt={The standard deviation of the effective temperature predictions increases at low and high temperatures, low surface gravities, low metallicities, high extinction, and extreme values of [N/O]. It scales with the residual on the effective temperature predication. The smallest values are on the order of a few 10 K, with the largest values being about 300 K, though most values are much smaller than that.}]{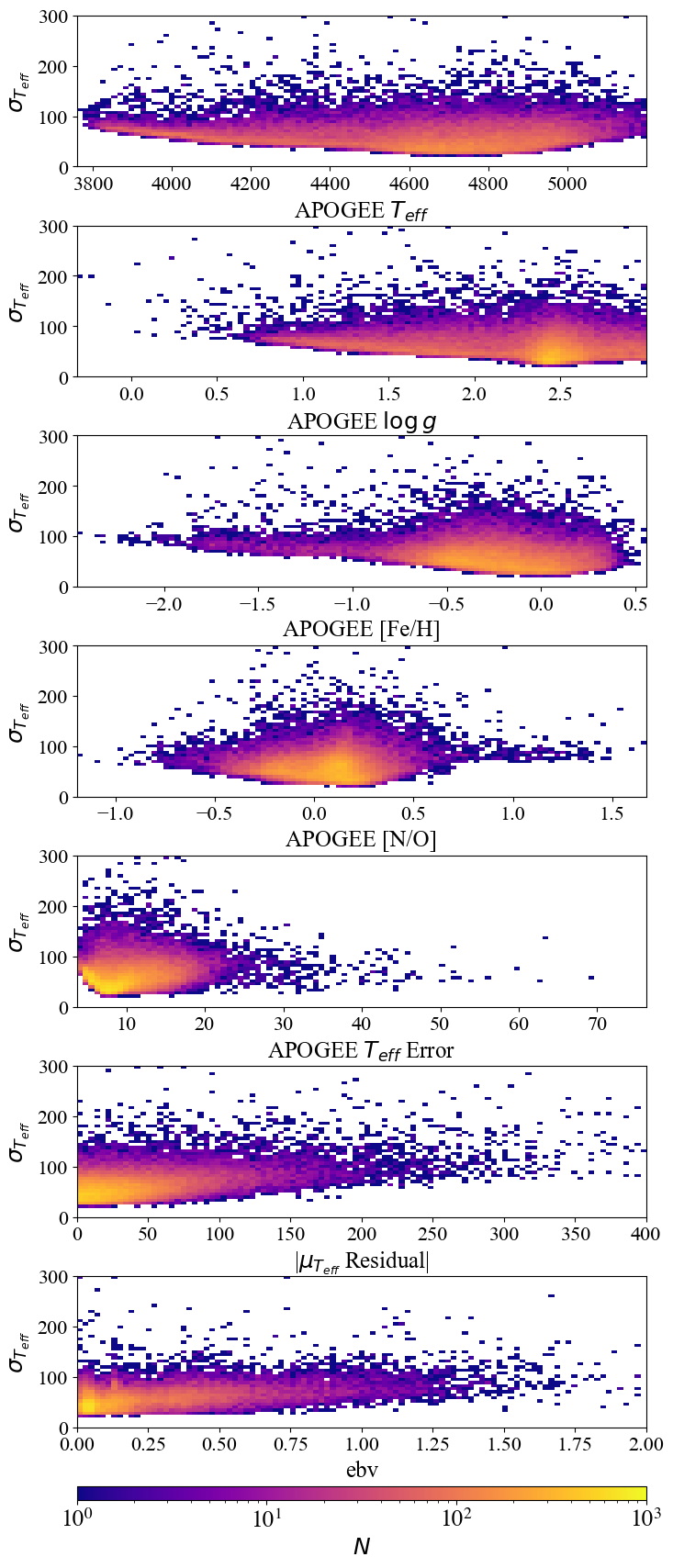}
    \caption{From top down, the MLP standard deviation prediction of $T_\mathrm{eff}$ ($\sigma_{T_\mathrm{eff}}<5000$) versus APOGEE values for $T_{\textrm{eff}}$, $\log g$, [Fe/H], [N/O], the residual of the $\mu_{T_\mathrm{eff}}$ prediction in the validation dataset, and $\textrm{E}(\textrm{B}-\textrm{V})$ from \citet{sfd_ebv}. The colorbar is shared for all panels and marks the color-mapping of the histograms as the log-scaled number of stars per pixel.}
    \label{fig:teff_stdev}
\end{figure}

\begin{figure}
	\includegraphics[width=\columnwidth, alt={The standard deviation of the log g prediction follows similar trends to the effective temperature standard deviation, with the standard deviation increasing at low surface gravities, low temperatures, high extinctions, and extreme values of [N/O]. The standard deviation again scales with the residual of the surface gravity prediction. Typical values for the standard deviation are below 0.5.}]{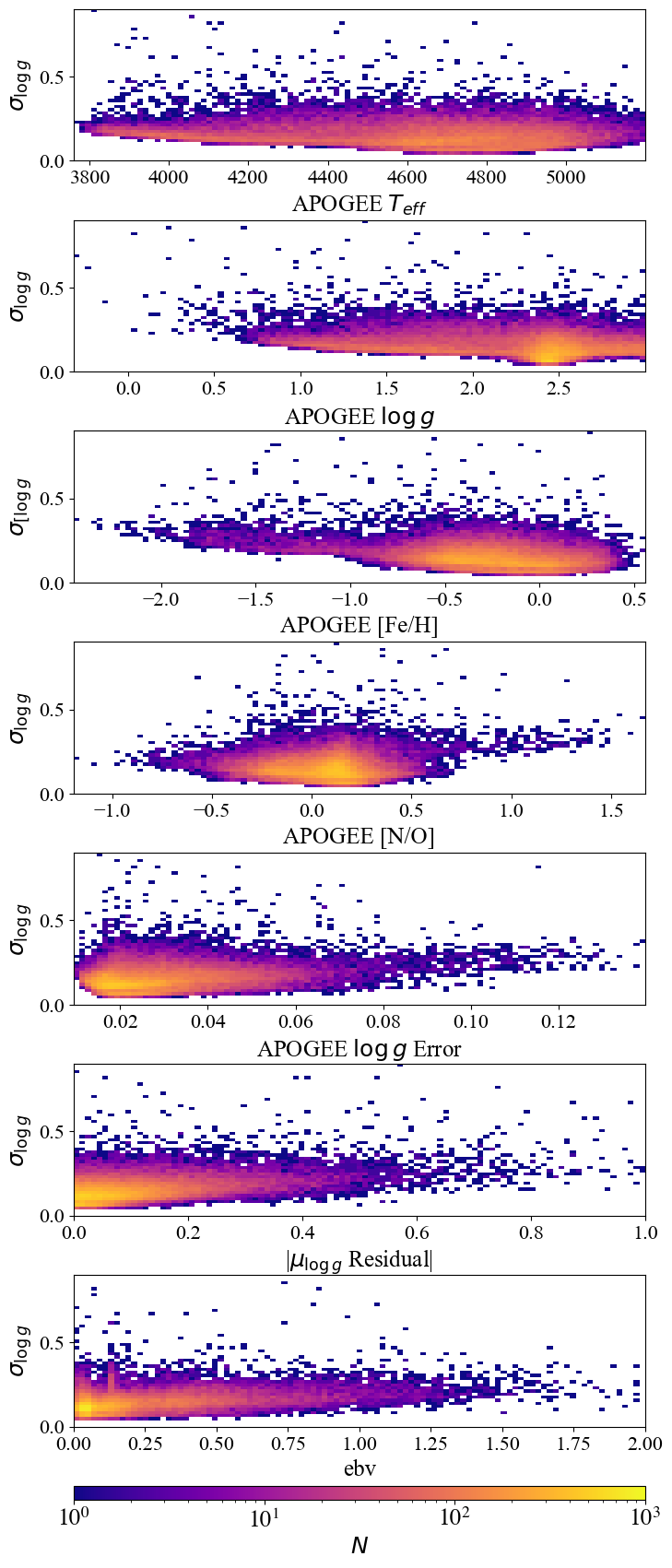}
    \caption{From top down, the MLP standard deviation prediction of $\log g$ ($\sigma_{\log g}$) versus APOGEE values for $T_{\textrm{eff}}$, $\log g$, [Fe/H], [N/O], the residual of the $\mu_{\log g}$ prediction in the validation dataset, and $\textrm{E}(\textrm{B}-\textrm{V})$ from \citet{sfd_ebv}. The colorbar is shared for all panels and marks the color-mapping of the histograms as the log-scaled number of stars per pixel.}
    \label{fig:logg_stdev}
\end{figure}

\begin{figure}
	\includegraphics[width=\columnwidth, alt={The standard deviation of the [Fe/H] prediction most clearly increases with decreasing metallicity and correlates weakly with the residual on the [Fe/H] prediction. The standard deviations are also high at high [N/O] values. Typical values for the [Fe/H] standard deviation are below 0.5.}]{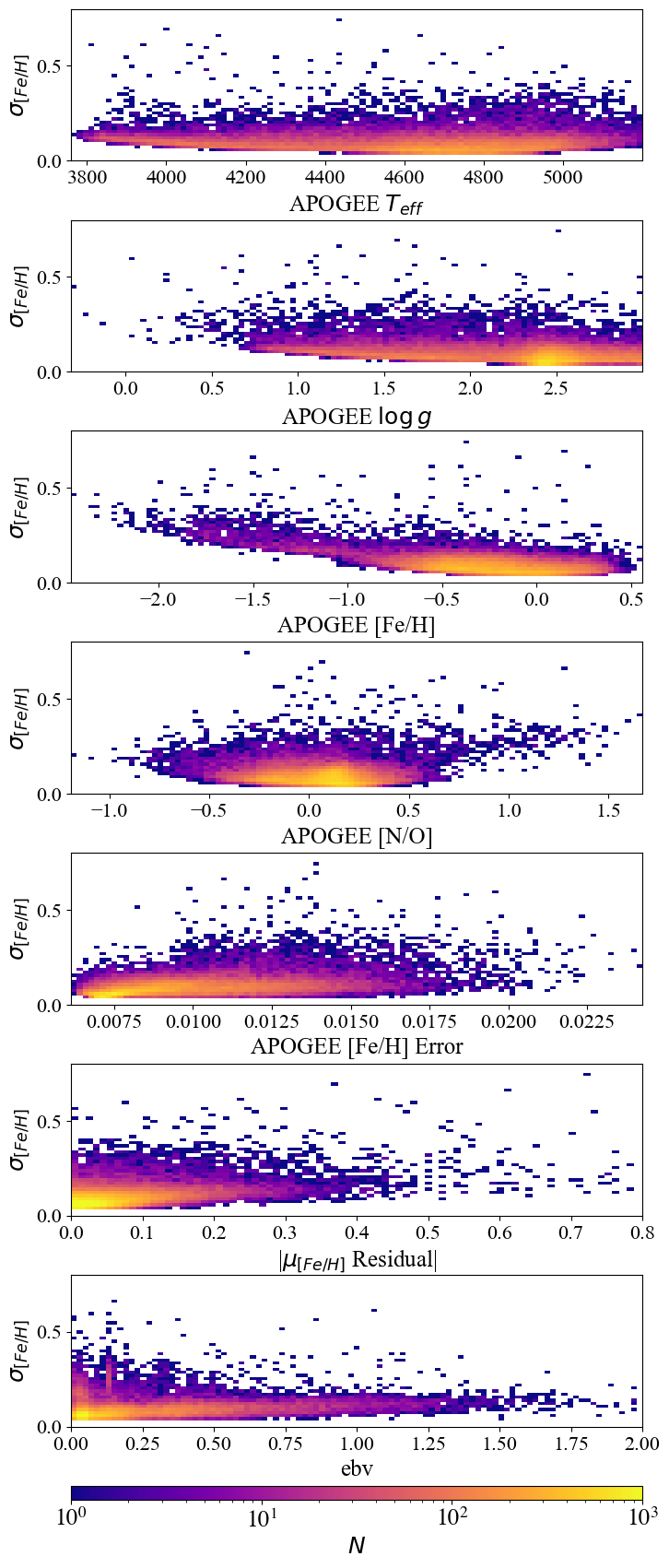}
    \caption{From top down, the MLP standard deviation prediction of [Fe/H] ($\sigma_{\textrm{[Fe/H]}}$) versus APOGEE values for $T_{\textrm{eff}}$, $\log g$, [Fe/H], [N/O], the residual of the $\mu_{\textrm{[Fe/H]}}$ prediction in the validation dataset, and $\textrm{E}(\textrm{B}-\textrm{V})$ from \citet{sfd_ebv}. The colorbar is shared for all panels and marks the color-mapping of the histograms as the log-scaled number of stars per pixel.}
    \label{fig:feh_stdev}
\end{figure}

\begin{figure}
	\includegraphics[width=\columnwidth, alt={The standard deviation of the [Al/Fe] prediction is highest at the extreme high values of [Al/Fe] and [N/O] and also increases with decreasing metallicity. There is no or only a very weak correlation with the residual on the [Al/Fe] prediction. Typical values for the standard deviation appear to be below 0.4 or 0.5.}]{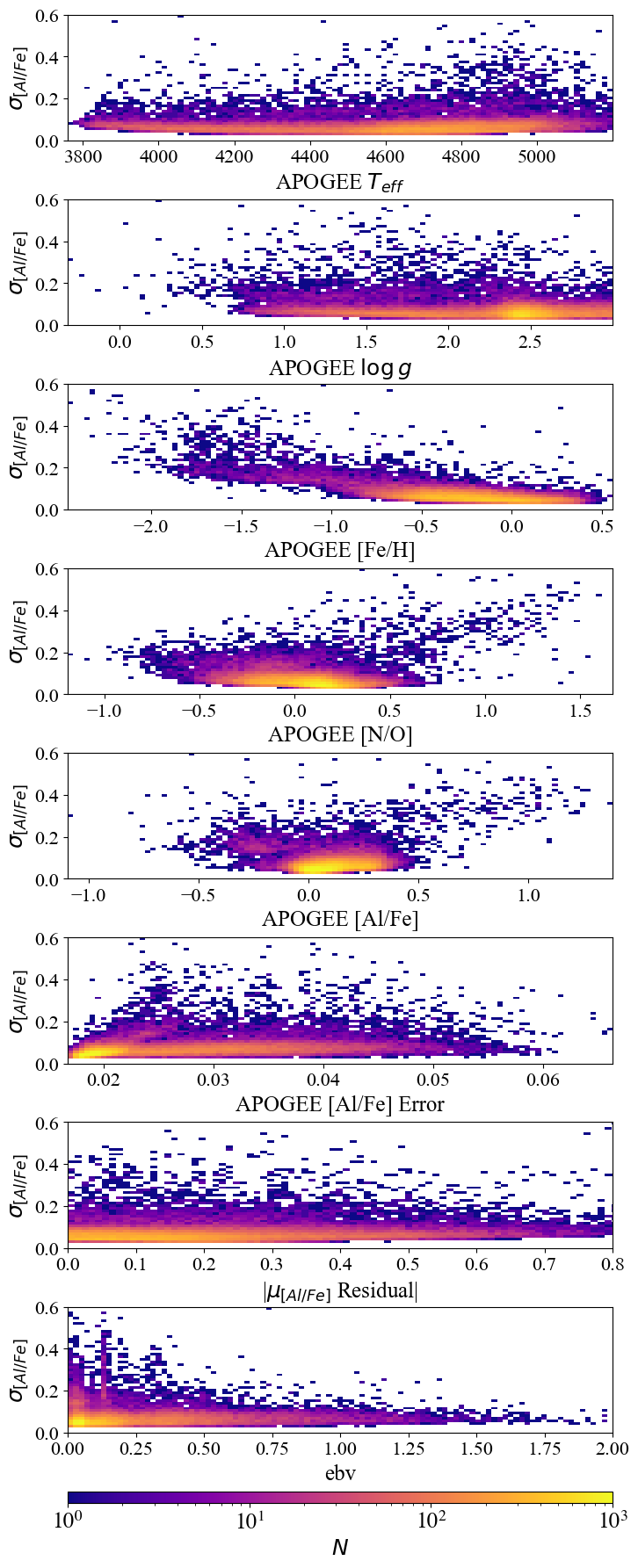}
    \caption{From top down, the MLP standard deviation prediction of [Al/Fe] ($\sigma_{\textrm{[Al/Fe]}}$) versus APOGEE values for $T_{\textrm{eff}}$, $\log g$, [Fe/H], [N/O], [Al/Fe], the residual of the $\mu_{\textrm{[Al/Fe]}}$ prediction in the validation dataset, and $\textrm{E}(\textrm{B}-\textrm{V})$ from \citet{sfd_ebv} The colorbar is shared for all panels and marks the color-mapping of the histograms as the log-scaled number of stars per pixel.}
    \label{fig:alfe_stdev}
\end{figure}


\bsp	
\label{lastpage}
\end{document}